\documentclass[manuscript,screen]{acmart}
\AtBeginDocument{%
  \providecommand\BibTeX{{%
    \normalfont B\kern-0.5em{\scshape i\kern-0.25em b}\kern-0.8em\TeX}}}
\usepackage{subfigure}
\usepackage{epigraph} 
\usepackage{graphicx}
\usepackage{tabularx}
\usepackage{array}
\usepackage{pdflscape}

\setcopyright{rightsretained}
\acmDOI{XXXXXXX.XXXXXXX}

\acmISBN{978-1-4503-XXXX-X/18/06}

\begin{document}

\title{\textit{What Do I Need to Design for Co-Design?} \newline Supporting Co-design as a Designerly Practice}


\author{Shruthi Sai Chivukula}
\email{schivuku@iu.edu}
\affiliation{%
  \institution{Indiana University}
  \streetaddress{901 E 10th St Informatics West}
  \city{Bloomington}
  \state{Indiana}
  \country{USA}}
  
\author{Colin M. Gray}
\email{gray42@purdue.edu}
\affiliation{%
  \institution{Purdue University}
  \streetaddress{401 N Grant Street}
  \city{West Lafayette}
  \state{Indiana}
  \country{USA}
  \postcode{47907}
}

\renewcommand{\shortauthors}{Chivukula \& Gray}

\begin{abstract} 
Co-design practices have been used for decades to support participatory engagement in design work. However, despite a wide range of materials that describe the design and commitments of numerous co-design experiences, few descriptions of the knowledge that guides designers when creating these experiences exist. Thus, we ask: What kind of knowledge do designers need to design co-design experiences? What form(s) could intermediate-level knowledge for co-design take? To answer these questions, we adopted a co/auto-ethnographic and Research-through-Design approach to reflexively engage with our design decisions, outcomes, and challenges related to two virtual co-design workshops. We constructed a set of four multi-dimensional \textit{facets} (Rhythms of Engagement, Material Engagement, Ludic Engagement, and Conceptual Achievement) and three roles (designer, researcher, facilitator) to consider when creating co-design experiences. We illustrate these facets and roles through examples, building new \textit{intermediate-level knowledge} to support future co-design research and design, framing co-design as a designerly practice.


\end{abstract}




\maketitle

{\color{red}\textbf{Draft: September 13, 2022}}

\section{Introduction}
Over the past two decades, co-design\footnote{Scholarship relating to generative and participatory experiences frequently uses the terms ``co-design,'' ``co-creation,'' and ``participatory design.'' Our goal in this paper is not to delimit any one of these terms, and instead we have standardized on the term ``co-design.'' We believe that our core knowledge contribution has the potential to advance all three common framings of generative and participatory engagement.} has become a common participatory approach and form of inquiry in the HCI community. This methodology has been used to encourage broader participation by involved stakeholders~\cite{Olivier2015-dt,Yoo2013-qu}, engage researchers in action-oriented research approaches that involve intentional change through design~\cite{Lee2008-sn,Hayes2014-mp}, support designerly inquiry practices through design-oriented empirical practices using framings such as Research through Design~\cite{Zimmerman2007-qi,Zimmerman2010-lm}, and consider the ``everyday'' role that design might play when engaging partners (often not professionally trained in design) in design work~\cite{Manzini2015-om}. 

\begin{figure}[hbt]
    \centering
    \includegraphics[width=\textwidth]{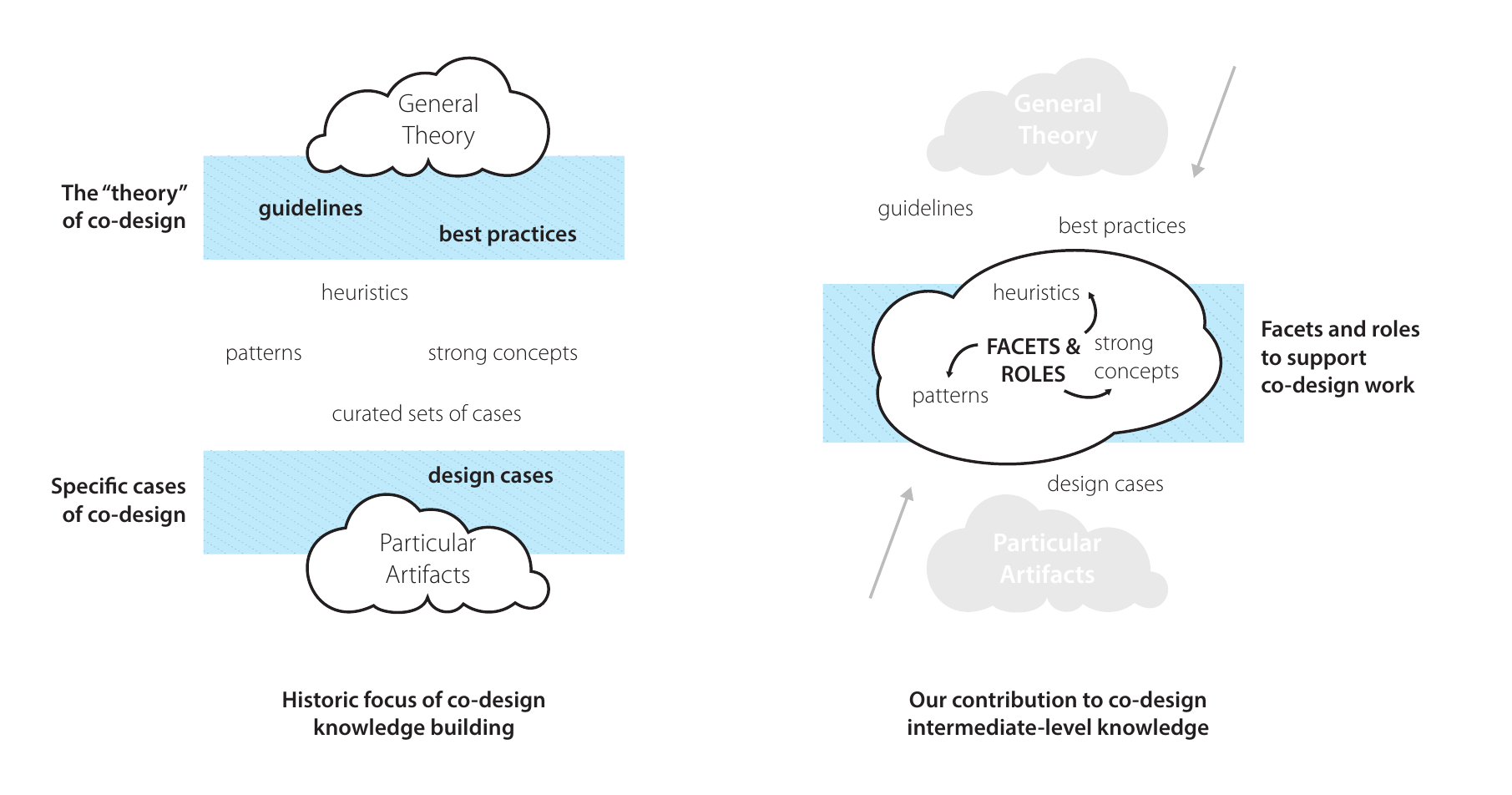}
    \caption{Historic intermediate-level knowledge creation in co-design scholarship compared with our desired contribution in this paper, building on a diagram from L\"owgren~\cite{Lowgren2013-db}.}
    \Description[Schema representing Intermediate-level knowledge from L\"owgren~\cite{Lowgren2013-db}]{Figure presents two schemas. The schema on the left is a representation of different types of intermediate-level knowledge between two big states called ``General Theory'' and ``Particular Artifacts.'' There are two highlighted areas (in Blue rectangles) in this schema representing the focus of co-design literature till date that focused more towards these end states. The schema on the right is a re-creation of the previous schema to highlight the areas of focus for this paper, which is represented by a blue rectangle in the area between the two end states. In this middle area, the figure lists ``Facets & Roles'' in big text drawing from ``heuristics,'' ``patterns,'' and ``strong concepts.''}
    \label{fig:ILK}
    
\end{figure}

Recent scholarship has demonstrated interest in describing the roles that facilitators of participatory experiences take on~\cite{Dahl2022-fg}, building on historic interest by participatory design (PD), co-creation, and co-design scholars in designing engaging experiences that empower and enable stakeholders (e.g., ~\cite{Lee2008-sn,Sanders2008-eq,Sanders2012-om,McKercher2020-xq}). While the literature to motivate generative and participation-driven approaches such as co-design is robust, there have been few accounts that explicitly describe the range of knowledge that the designers of co-design experiences rely upon when making key design decisions, or contributions to knowledge that do not primarily seek to advance theory-building or generation of principles to guide future co-design engagement (as highlighted in Figure~\ref{fig:ILK}, left). In this paper, we seek to build new \textit{intermediate-level knowledge} to support future co-design work, explicitly framing the design of co-design experiences as a designerly practice. By intermediate-level knowledge, we refer to the ``\textit{constructi[on of] knowledge that is more abstracted than particular instances, yet does not aspire to the generality of a theory}''\cite[23:2]{Hook2012-dd}. Like H\"o\"ok and L\"owgren, our goal is to support generative design practices, and in the case of co-design practices, in this paper we build new knowledge that supports the designer in making judgments that allow them to create new co-design experiences.

We use a co/auto-ethnographic approach~\cite{Coia2009-bf}, framed through Research through Design engagement that prioritizes the practice of the designer as a form of legitimate knowledge generation~\cite{Zimmerman2007-qi}. We report on an 18--month design investigation that resulted in the creation of two digital co-design experiences, leveraging our reflection on these design experiences to form a set multi-dimensional set of \textit{four facets} and \textit{three ``hats'' or roles} that we considered when engaging in design work. The four facets include: 1) \textit{Rhythm of Engagement} that considers designing over space and time in a co-design space; 2) \textit{Material Engagement} that considers designing of physical co-design material and logistics; 3) \textit{Ludic Engagement} that considers designing aesthetic experiential aspects of a co-design space; and 4) \textit{Conceptual Achievement} that considers designing for goals to be achieve through co-design. Considering these facets, the three roles include \textit{Designers} and \textit{Researchers} who are involved in designing the co-design space and material, and \textit{Facilitators} who are involved in on-the-ground engagement and directing in a co-design workshop. Through these facets and roles, we build new intermediate-level knowledge (as highlighted in Figure~\ref{fig:ILK}, right) to support the design of future co-design experiences and also lay the groundwork for new kinds of knowledge building that exist between the endpoints of theory and specific case studies (often-explored in literature as shown in Figure~\ref{fig:ILK}, left).

In this paper, we make two primary contributions. First, we interrogate our own co-design practices through a co/auto-ethnographic and Research through Design lens, framing the design of co-design as a designerly practice by laying the groundwork for how key design decisions are made when designing for co-design and describing details of two different digital co-design experiences. Second, we articulate an initial set of intermediary-level knowledge, framed through four key facets and three roles, that could support the description of design decisions that shape the design of co-design experiences and scaffold future co-design practices. 

\section{Background Work}
To frame our desired contribution, we first provide a foundation of literature that we relied upon when engaging in \textit{co-design work}, demonstrating the range of knowledge that researchers and practitioners engage in and with when creating new co-design experiences. We then frame this knowledge as part of a larger historical investigation into the kinds of \textit{design knowledge} that designers rely upon to support their work, identifying areas where we anticipate building new knowledge as a contribution of this paper. 

\subsection{Co-Design as a Methodology to Encourage Participation}

Participatory approaches in design and HCI and contexts have evolved over the last three decades, marked by a shift from user-centered design processes and methods, with a typical focus on the designed artifacts reaching the goals or needs of the user, to a participatory culture that seeks to enable and empower diverse stakeholder participation in design work~\cite{Sanders2002-na}. Sanders and Stappers \cite{Sanders2008-eq} define co-creation and co-design as a part of this shift towards participatory approaches, which defines new roles and opportunities for users to take on. In a broad sense, co-creation is defined as ``any act of collective creativity, i.e. creativity that is shared by two or more people'' and co-design is defined as ``collective creativity as it is applied across the whole span of a design process'' \cite[p. 6]{Sanders2008-eq}. There is often confusion between these terms, but these concepts can be differentiated based on who is involved in the act, the goals of involvement, and when they are involved in the design process \cite{Marttila2013-ax}. Co-creation has its origins in the fields of business studies and marketing, whereas co-design draws more from design-focused fields following participatory design or co-operative design traditions \cite{Durall2020-zz}. These two terms are often intertwined, but building on the definitions, we consider co-design as an instance of co-creation, and standardize on the term co-design throughout this paper as our point of focus. In the following sub-sections, we will provide a broad outline of knowledge that has been generated in relation to co-design practices and experiences, using the left side of Figure~\ref{fig:ILK} to indicate knowledge-building that has been focused on theory-building or generation of principles to guide future co-design engagement (Section~\ref{sec:codesignguidelines}) and knowledge-building that has been focused on specific examples of co-design environments (Section~\ref{sec:codesigncasestudies}). 

\subsubsection{Guidelines and Principles} \label{sec:codesignguidelines}
Co-design is an approach that seeks to treat ``users as a partner,'' giving the designers or researchers access to the tacit knowledge of the users as they participate in the design process \cite{Sanders2008-eq}. Co-design also has many similarities or principles that align with contemporary participatory approaches like design justice, which focuses on principles such as: ``center the voices of those who are directly impacted,'' ``everyone is an expert based on their lived experience,'' and ``designer as a facilitator than an expert'' \cite[p. 6]{Costanza-Chock2020-vf}. However, co-design is not always framed in ways that foreground politics and does not always begin with the goal of empowering a set of users. In a co-design process, there is often involvement of diverse actors such as researchers, designers or developers, and users (often framed as citizens with approaches such as digital civics~\cite{Olivier2015-dt}). Co-design participants are treated as having expertise about their own experiences, and are ideally given agency to support the activation of this knowledge \cite{Sanders2008-eq,Steen2011-qo,Visser2005-kl}. 

Scholars of co-design have placed a focus on differing aspects of these experiences which prioritize certain qualities while potentially deprioritizing others. These areas of focus are frequently framed as principles, best practices, or other contributions to a theory of co-design as a legitimate form of inquiry. For instance, \textit{embodiment} is a described as a typical goal of well-designed co-creation experiences. Kronqvist and Salmi~\cite{Kronqvist2011-py} note multiple qualities of this potential embodiment, including: ``\textit{both the physical and social aspects of engagement with the world [\ldots through which] participants collaboratively envision the future through interacting with each other and with physical materials creating prototypes, models, sketches, collages, posters, stories to name a few examples}.'' Similarly, other scholars have described co-design as containing ``an embodied continuum'' \cite{Akama2013-pb,Light2012-zg} rather than a process that is solely object-focused. Another dominant theme is co-design as a site for \textit{shared meaning-making}, which Kleinsmann and Valkenberg~\cite{Kleinsmann2008-du} define as a ``process in which actors from different disciplines share their knowledge about both the design process and the design content [\ldots] in order to create shared understanding on both aspects [\ldots] and to achieve the larger common objective: the new product to be designed.'' Building upon both the object orientation of co-design and its focus on building shared meaning making, Evans and Terrey \cite{Evans2016-lr} also frame these practices in ways that foreground \textit{reflection}, where participants engage in ``a methodology of research and professional reflection that supports inclusive problem solving and seeks solutions that will work for people.'' Finally, co-design scholars have described these experiences as being focused on \textit{collaborative and generative engagement}. Steen \cite{Steen2013-is} describes the collaborative dimension of co-design as a ``\textit{process of joint inquiry and imagination in which diverse people jointly explore and define a problem and jointly develop and evaluate solutions;}'' whereas, Sanders and Stappers \cite{Sanders2012-om} focus attention on the generative thrust of the method with valuable outcomes for designers as a design research method, framing co-design as ``an approach to bring the people we serve through design directly into the design process to ensure that we can meet their needs and dreams for the future.'' 

Other frameworks have been proposed to describe a linked set of commitments that are germane to co-creation work, exceeding any one of the principles described above. In a recent book, McKercher \cite{McKercher2020-xq} defined four principles of co-design as follows: share power, prioritize relationships, use participatory means, and build capacity. Lee and colleagues \cite{Lee2018-ne} suggested a framework to guide co-design work which emerged from a cross-case analysis of 13 co-design projects, which consists of ten design choices facilitators have to make, presented in four categories: participants (diversity in knowledge, differences in interest, distribution of power), project preconditions (openness of the brief, purpose of change, the scope of design), co-design events (types of activities, setting for co-creation), and project results (outputs of the project, outcomes of the project). McKercher \cite{McKercher2020-xq} defined six mindsets for co-design, namely: Elevating lived experience, Practising curiosity, Offering hospitality, Being in the grey, Learning through doing, and Valuing many perspectives.  

\subsubsection{Case Studies}
\label{sec:codesigncasestudies}
Co-design has been deployed as a method for inquiry and user engagement across a range of domains, including health services \cite{Cottam2004-di,Kuo2019-uh,Shakespeare2014-rf}, transport services \cite{Yoo2010-gy}, organizational changes \cite{Kronqvist2011-py,Madaio2020-gu,Steen2011-ue,Yoo2013-qu}, public space or urban designs \cite{Cruickshank2013-af,Dede2012-mh,Del_Gaudio2018-zi}, mobile or digital interactions \cite{Fitton2018-cw,Loi2019-tl,Stigberg2017-ni,Yoo2013-qu}, policy design \cite{Kuo2019-uh,Madaio2020-gu}, and game design \cite{Vaajakallio2014-bu}. Across these examples of co-design projects in the literature, a co-design methodology was selected and primarily used to support two main purposes. First, co-design was used to build new, innovative and creative solutions. For example, projects included generating ideas for new Telecom Services with employee’s children \cite{Steen2011-ue}, the development of ``context cards'' to improve bus-transportation with everyday bus transport users \cite{Hilden2017-jk}, and the design of network services for the homeless with young people experiencing homelessness, and police and service providers \cite{Yoo2013-qu}. Second, co-design was used to change existing forms of design or contexts. For example, project partners engaged in co-design to conduct internal workshops with employees to improve current logistic services for customers \cite{Steen2011-qo}, support a large scale co-designing of a park with citizens beyond the existing castle \cite{Cruickshank2013-af}, build an organizational level effort to impact change by taking inputs from the employees for new innovation practices \cite{Kronqvist2011-py}, and spearhead a service-focused co-design effort for a better pharmacy service with the pharmacy users \cite{Shakespeare2014-rf}. Across this range of published work, researchers and designers presented their co-design work in the form of illustrative case-studies that was used as a value-added approach suitable to support their research design; a methodology to illustrate discovery, innovation, and insight that involved the end-customers in the process; and very rarely, a means of discussing how they engaged in the design of their co-design material and spaces. This latter purpose falls under the knowledge close to that of specific precedent examples that is translated into particular artifacts through the application of various co-design principles and mindsets, which we will discuss further in the next section.  

\subsection{Designerly Practice and Design Knowledge}

The study of design cognition and designerly ways of knowing has a long history---beginning with the birth of the design methods movement in the 1960s and continuing to the present day in both discipline-specific and discipline-agnostic forms~\cite{Cross2001-zg,Cross2007-ei}. In the decades since, design theorists have focused their attention on numerous aspects of designerly practice, including the capacity of the designer themself~\cite{Schon1990-by,Boling2020-ci,Nelson2012-ov}, the sources of knowledge that designers draw upon~\cite{Lawson2004-wf,Kolaric2020-qc,Hook2015-tt}, the ecological setting in which design work occurs~\cite{Stolterman2008-ho,Goodman2011-ak,Gray2019-ep}, and the situated judgments they make to inform design decisions~\cite{Nelson2012-ov,Dunne1999-ub,Gray2015-qi,Gray2017-dx,Christensen2017-tm}---all of which inform the designed outcomes that result from their engagement~(c.f., \cite{Schon1983-dl,Verbeek2006-qz}). 

Notably, the work of a designer is complex and situated. Stolterman~\cite{Stolterman2008-ho} provides one useful description of the multifarious nature of design activity through the concept of design complexity. Design complexity refers to ``the complexity a designer experiences when faced with a design situation [\ldots including] mak[ing] all kinds of decisions and judgments, such as, how to frame the situation, who to listen to, what to pay attention to, what to dismiss, and how to explore, extract, recognize, and chose useful information from all of these potential sources''~\cite[p. 57]{Stolterman2008-ho}. Engagement with this design complexity is inevitable---critically involving a designer's negotiation with design knowledge in many forms, and particularly of importance in fields of design that are actively being shaped by practitioners, educators, and researchers~\cite{Kuo2019-uh}.

Design theorists have identified many different forms of design knowledge, with Lawson~\cite{Lawson2004-wf} describing schemata, design gambits, and design precedent as three different levels or layers of knowledge, Sch\"on~\cite{Schon1990-by} describing designer's engagement with a repertoire of design precedent materials, Boling and Smith~\cite{Boling2012-od} in describing how designers might productively build precedent knowledge, and Gray et al.~\cite{Gray2016-lq} extending Sch\"on's notion of repertoire to include knowledge of methods as well as precedent artifacts. L\"owgren and colleagues ~\cite{Lowgren2006-vr,Lowgren2013-db} have also identified a taxonomy of design knowledge that describes two different ``poles'' of knowledge forms that relate to theoretical knowledge and precedent knowledge with a large space identified in between these poles, which they describe as \textit{intermediate level knowledge}~\cite{Hook2012-dd,Lowgren2013-db}~(see also Figure~\ref{fig:ILK}, left). Many kinds of knowledge are claimed to be contained in this middle space, including use qualities~\cite{Lowgren2006-vr}, annotated portfolios~\cite{Lowgren2013-db}, patterns~\cite{Tidwell2010-tq},  strong concepts~\cite{Hook2012-dd}, and precedent artifacts~\cite{Lawson2004-wf,Boling2012-od}. We build upon this schema of knowledge production, focusing our attention upon building knowledge in the center part of the diagram~(Figure~\ref{fig:ILK}, right).

In this paper, we frame our engagement in designerly practice, particularly drawing on Sch\"on's notion of reflective practice \cite{Schon1983-dl} in a pragmatist framing to describe how the designers of co-design experiences rely upon and build knowledge through their engagement in design work. In particular, we seek to add further clarity to the formation of intermediate-level knowledge that may support co-design work, augmenting existing interest in documenting principles, best practices, and case studies of co-design experiences with the identification of a set of four facets and three roles which we describe in Sections~\ref{sec:facets} and \ref{sec:hats}.


\section{Our Approach}
We followed a co/autoethnography approach \cite{Coia2009-bf}, a method inspired from the education research literature used to ``\textit{examine [our] teaching selves.}'' Using this approach as a launch-off point for inquiry, we (as the two authors of this paper) collaboratively documented and shared our experiences as we ``examined'' our designer selves through the design and implementation of two different co-design workshops (detailed in Section \ref{CaseA} and \ref{CaseB}). As part of this design work, we sought to engage in Research through Design (RtD) \cite{Zimmerman2007-qi,Zimmerman2014-zx}, relying upon our knowledge of and prior experience in leveraging participatory design principles \cite{Schuler1993-ut} and co-design methodologies \cite{Gaver1999-fi,Sanders2008-eq, McKercher2020-xq}. 
We used co/autoethnography as a reflexive tool for us to engage in structured conversation throughout the design and implementation processes, concretizing outcomes of this RtD work which happened at the intersection of our design practice and research engagement, which are detailed in the later sections of the paper. This reflexive process comprised memoing design process moves, failures, or iterations; sharing and de-briefing designer decisions; reflection on our experiences and sources of inspiration or sense-making; recording and documenting group discussions with other designers involved regarding ways to improve our co-design workshops; and photo-elicitation through design production. This wide range of communication served as an audit trail of artifacts that we used to frame the findings of this paper, including hand-written notes, video/audio recordings, whiteboard pictures, scans of sketches, transcripts of discussions, living documents of design decisions, and screenshots of Slack conversations.     

\subsubsection{Different Hats}
During the design process, we intentionally wore ``different hats'' as a part of our RtD approach \cite{Zimmerman2007-qi}, including as \textit{designers} of the co-design materials and spaces; \textit{researchers} in the domain of ethics in HCI and design to accomplish research goals associated with the co-design workshops; \textit{facilitators} who anticipated conducting these co-design workshops with participants; and \textit{RtD practitioners} who reflexively built upon the knowledge bases of participatory approach principles and our practical knowledge built through planning and conducting these workshops to create the intermediate knowledge presented in this paper. This juggling of these different hats is exemplified by a quote from the second author to the first author early in the design process, which described our goals of reflexive designerly practice: ``\textit{I think you're a designer. You're a designer that's designing with the intent of facilitation. But you're not facilitating yet.}'' This highlighted the need to define ``Hats'' for various roles that are involved in designing for co-design (in Section~\ref{sec:hats}); that we provide as a lens one has to take while engaging with four primary Facets (Section~\ref{sec:facets}) listed in this paper.


\begin{figure}
    \centering
    \includegraphics[width=0.95\textwidth]{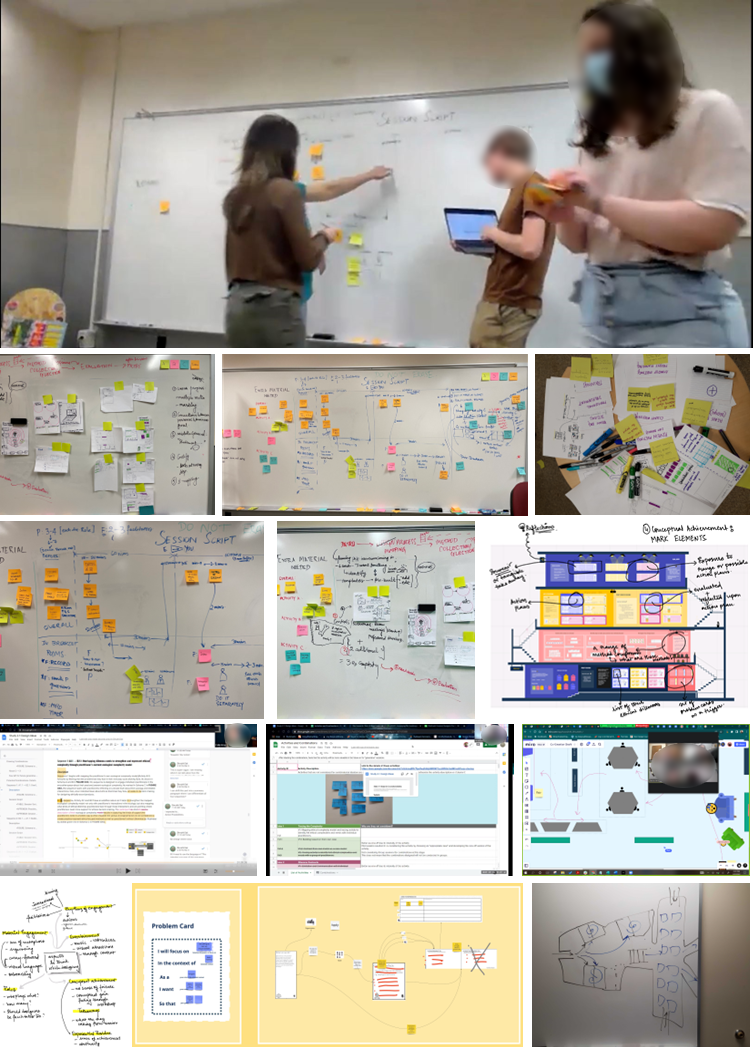}
    \caption{Examples of prototypes and other artifacts from our collaborative design of the co-design environment and materials.}
    \label{fig:collage}
    \Description[Figure consisting multiple photographs]{Figure consists of a collage of a variety of photographs captured by the designers during the design process of Case A and Case B co-design workshops detailed in this paper. The photographs include a group of designers working on a whiteboard, multiple images of whiteboards captured in the process, Zoom screenshots of collaborative working and editing on Miro board or Google Docs, and screenshots of some design medium-level prototypes created for ideation and creation of the final prototype.}
\end{figure}

\subsubsection{Our Experience in Designing for Co-Design}
The authors of this article hold degrees earned in design including HCI design, visual design, and instructional design, which was augmented by four or more years of engaging in ethics-related research that prioritized designerly approaches. We identify ourselves and locate our positionality as design researchers who have engaged in a range of research methods and types of design production which supported our inquiry. Author 1\footnote{Left blank for the purpose of anonymous review} has ten years of experience in UX design and Author 2 has over 20 years as a designer in instructional design, visual design, and UX design capacities. 
 
Beyond our individual expertise, our history of interactions as designers and researchers also frames this project. We have conducted ethics-related research in HCI for the past four years and have built an extensive landscape of knowledge related to ethics in HCI and design practice that supports our present analysis and design experiences\footnotemark[2]. In this paper, we rely upon two such cases of collaboration conducted over an 18--month period, through which we share our ``behind the scenes'' perspective and leverage these efforts to produce new design knowledge. In Case B, we co-led a design team of undergraduate and graduate researchers that contributed to our final desired research and design outcomes, with a collective research team goal of engaging technology and design practitioners to create bespoke methods that could support their ethical responsibility in their context of everyday work (workshop described in Section~\ref{CaseB}). In Case A, Author 1 designed co-design materials and piloted one-on-one workshops with design and technology practitioners as a part of her dissertation (workshop described in Section~\ref{CaseA}). 

In reflecting on both of these experiences as designers in the moment, and in framing this work in the tradition of RtD, we were able to not only design co-design workshops together which were effective for their intended purpose, but also leveraged our designerly practice across the entire process with a range of design decisions that led to the creation of new design knowledge. Additionally, given that these co-design efforts took place during the pandemic, we were constrainted in using a \textit{digital medium} to conduct the workshops. This less typical medium for co-design work, while not entirely new but rare, helped us to defamiliarize ourselves and allowed us to question our utilization of prior design precedent that primarily stemmed from physical co-design experiences.

\section{Case Studies: Co-design Workshop Examples} \label{sec:casestudydescriptions}
In this section, we describe two cases of co-design workshops we as co-authors designed together and reflected upon to construct the facets of designing for co-design. These workshops were created by a team of more than five designers that we co-led, including both graduate and undergraduate students trained in design and qualitative research methods. The designing of the co-design experiences was a team effort, but as co-authors, we led the team in reflection about key design decisions and aspects of the design process. 

\subsection{Case A: Engaging Practitioners through Co-Creation to Reflect on Ethics in Technology Practice} \label{CaseA}

\begin{figure}[ht]
    \centering
    \subfigure[]{\includegraphics[width=0.45\textwidth]{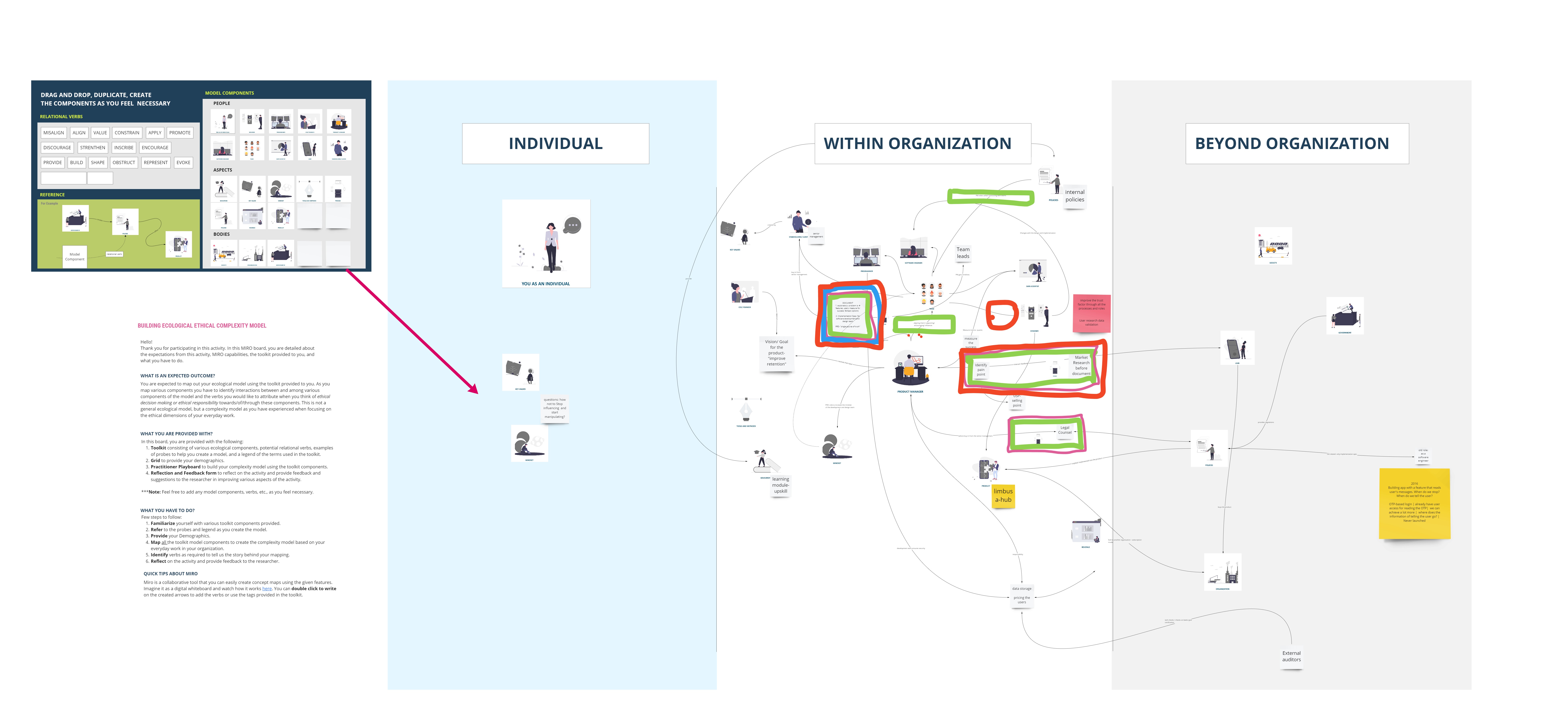}} 
    \subfigure[]{\includegraphics[width=0.45\textwidth]{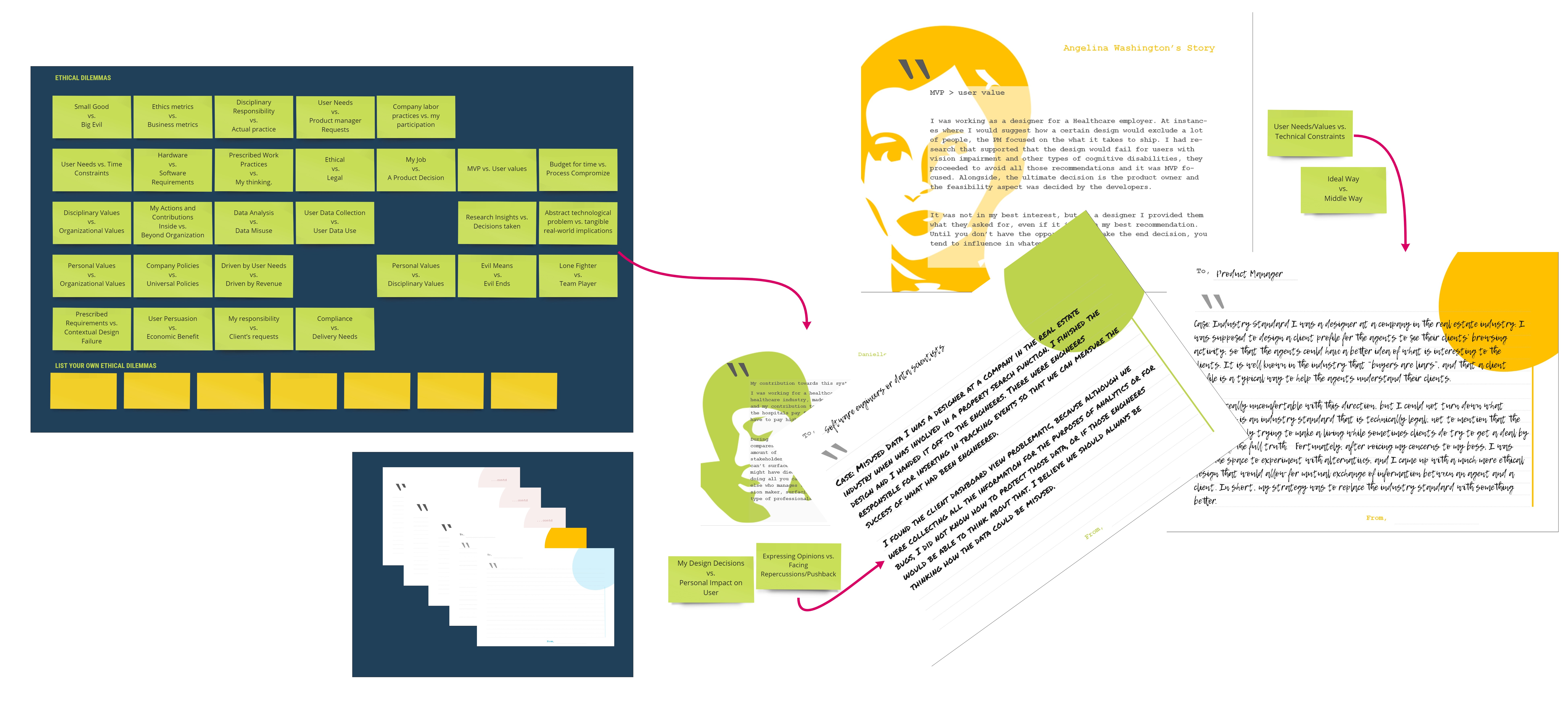}} 
    \subfigure[]{\includegraphics[width=0.5\textwidth]{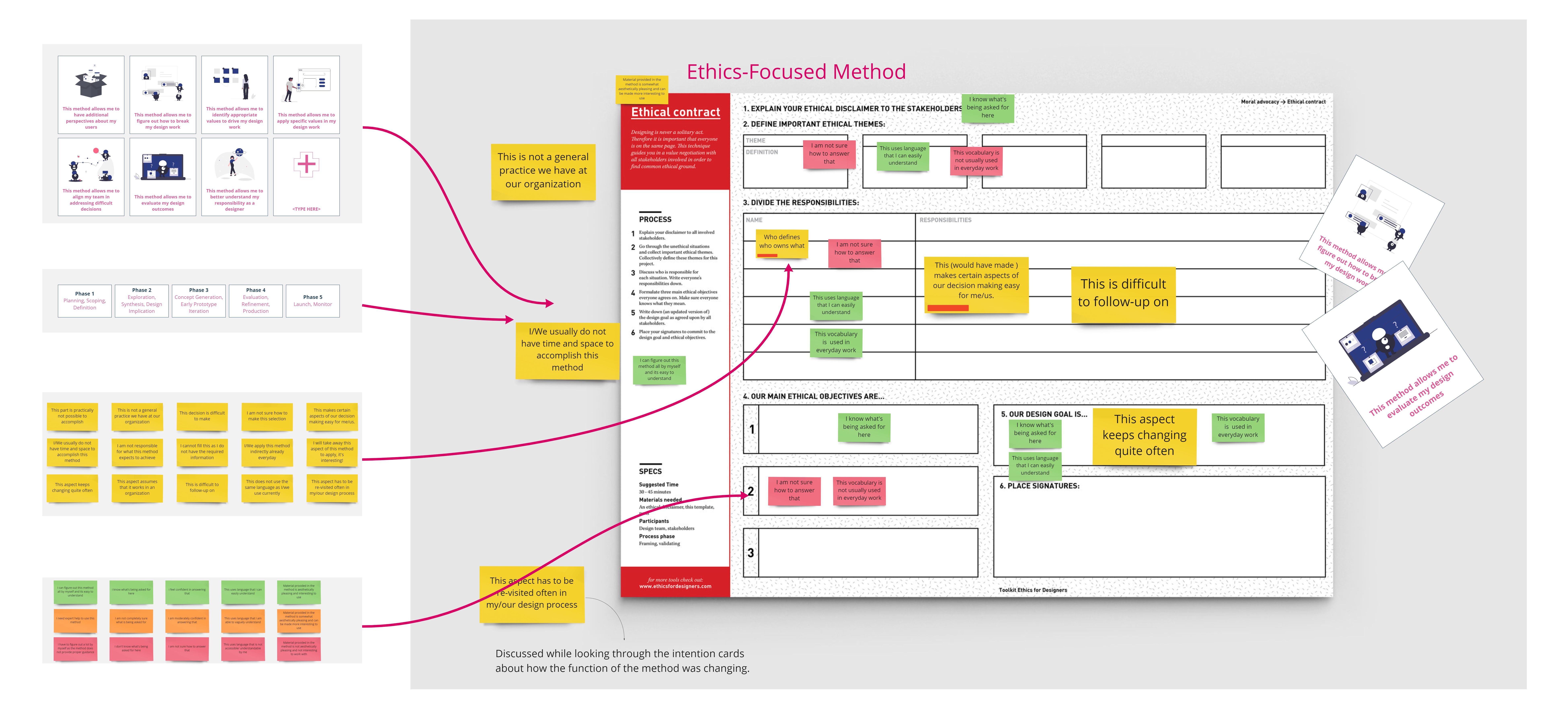}}
    \caption{(a) Activity A: Tracing the Complexity (b) Activity B: Filtering Ethical Dilemmas (c) Activity C: Evaluating using Method Heuristics}
    \Description[Figure with three sub-figures]{Figure consisting of three sub-figures named as (a), (b), and (c) of different activities described in Case A in the paper. Sub-figure (a) has a worksheet with a complex mapped ecological model with some image cards given to the participant. sub-figure (b) has a couple of filled postcards with ethical dilemma stories from technology and design practitioners. Sub-figure (c) has a toolkit and worksheet (which arrows identifying the connections between both) consisting of the method called Ethical Contract \cite{ethicalcontract} with some digital post-its placed by the participant over it as a part of the activity.}
    \label{fig:CaseA}
\end{figure}

\subsubsection{Our Goal} 
Our design goal for Case A was to design a one-on-one co-design space to facilitate a practitioner's engagement and reflection regarding ethical dimensions of their work. To accomplish this purpose, we designed three different co-design activities that consisted of interactive toolkits and probes for the participants to engage with. The primary goal of these activities was to engage and support practitioners in communicating about their felt ethical concerns in their primary professional role and as they interact with practitioners from other roles, with the goal of identifying the kinds of practices they seek to be better supported to engage or expand their primary role-focused notions of ethics and as they interact with practitioners from other professional roles. 

The design process of the co-design material and space in Case A was led by the first author over a period of three months. For each activity, the design process began with iteratively designing the main toolkit/probe as a paper prototype and then constructing the materials in high-fidelity on a Miro digital whiteboard. These materials were then framed through the design of facilitation materials for the co-design workshop with the participant, such as session scripts, probing questions, sequencing information, and Zoom setup. Finally, the first author conducted pilot sessions to iteratively improve the co-design material and space. 

\subsubsection{Co-design Material and Final Product}
In Case A (as shown in Figure \ref{fig:CaseA}), we designed three main activities to be experienced on an interactive Miro board. Each activity was expected to take 45-60 minutes for a participant to complete.
\begin{itemize}
    \item \textbf{Activity A---Tracing the Complexity:} A mapping activity for practitioners to create their ecological complexity model, sketching different people, aspects, and bodies involved and identifying the kinds of support they seek to support their ethical decision making in the mapped model. The Activity A toolkit was designed for participants to create their ecological model by \textit{mapping} and \textit{drawing connections} among a given set of components such as people, artifacts, and ecologies they interact with in their everyday work using a think-aloud protocol.
    
    \item \textbf{Activity B---Filtering Ethical Dilemmas:} A probe kit for practitioners to filter, reflect, elaborate, and speculate about ethical dilemmas they have faced in relation to their ethical commitments, decision making, and the complexity of their everyday work. The Activity B probe was designed for the participants to \textit{filter} a set of ethical dilemmas on a worksheet based on how they have either faced, not faced, faced in the past, or seen other face the given set of ethical dilemmas, and then \textit{narrate} stories around these dilemmas.
    
    \item \textbf{Activity C---Evaluating using Method Heuristics:} An evaluative activity for practitioners to evaluate, map, and re-imagine ethics-focused methods based on their resonance with their design activity and various ecological settings. The Activity C toolkit was designed for the participants to \textit{annotate} a provided ethics-focused method, \textit{Ethical Contract}~\cite{ethicalcontract}, using the provided heuristics designed in the form of small digital Post-Its to annotate the method on the Miro board. 
\end{itemize}

\subsection{Case B: Supporting Practitioner Co-Design of Action Plans to Activate Ethics in Their Everyday Technology Practice } \label{CaseB}

\subsubsection{Our Goal}
Our design goal for Case B was to design a co-design space to engage and facilitate a heteregeneous group of technology and design practitioners in designing an ethics-focused action plan (their expected final artifact) that they could possibly implement in the future to support their everyday work. For this purpose, we designed a sequence of different activities as a part of a co-design workshop that scaffolded participants' design of their action plan, which we explain in more detail in the following section. The primary goal of the entire step-by-step process was for participants to reflect on a range of ethical concerns in their everyday work, identify an ethical concern in a context that they wish to be better supported, choose from a range of existing ethics-focused tools as building blocks, create an action plan, and then evaluate the action plan to guide iterative improvements. 

\begin{figure}[ht]
    \centering
    \includegraphics[width=0.95\textwidth]{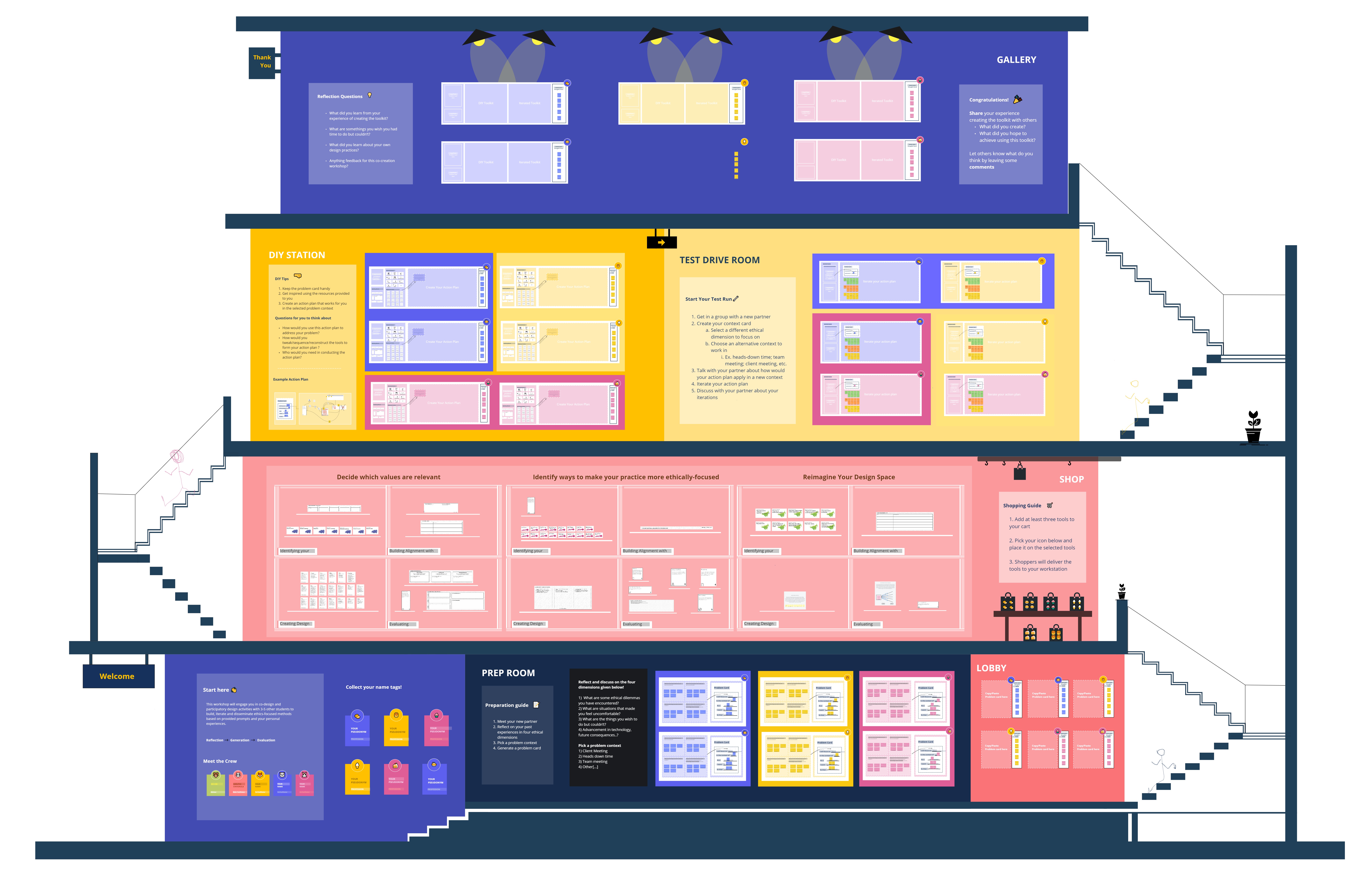}
    \caption{Case B---Co-design Material Design}
    \label{fig:CaseB}
    \Description[Schema of Case B in the form of a four-floored house]{Figure consists of co-design material designed for Case B described in this paper. The co-design material designed looks like a house with four floors. Each floor is designed for a particular design process as described in detail in Case B: Co-design Material and Final Product Section.}
\end{figure}

\subsubsection{Co-design Material and Final Product}
In Case B (as shown in Figure \ref{fig:CaseB}), we designed a 2-3 hour co-design workshop using a ``house'' metaphor with a range of activities that would guide the participants through building an ethics-focused action plan. The different stages, sequentially represented as \textit{rooms} in the visual metaphor in Figure \ref{fig:CaseB}, include:
\begin{itemize}
    \item \textbf{Welcome Room (Floor 1 West)---Introduction (15 mins):} This section of the workshop marks the on-boarding to the entire workshop where the facilitators and participants are given name tags to introduce themselves. This room includes material such as an introduction script, name tags, and specific identifiable icons and color for each participant representing the board they would be continuing to use throughout the workshop.  
    \item \textbf{Prep Room (Floor 1 Mid)---Scoping and Problem Identification (20 mins):} In a breakout room, this section of the workshop was designed for the participants to individually identify some of the ethical concerns they face in their everyday work using a probes and Post-Its. Based on their list of ethical concerns, they are asked to pick one of them they would like to build an action plan for by creating a ``Problem Card,'' which would carry throughout the length of the workshop. 
    \item \textbf{Lobby (Floor 1 East)---Presentation (5 mins):} As a group, this section of the workshop was designed for the participants to share their ``Problem Cards'' with other workshop participants to cross-pollinate other's perspective and allow for participants to leave any comments using the Post-Its. This section was also designed to be a part of a break in the session. 
    \item \textbf{Shop (Floor 2)---Collect Resources to Design (15 mins):} Individually, this section of the workshop was designed for the participants to collect provided method components derived from existing ethics-focused methods as building blocks to start composing their action plan in the next stage. The participants are given a chance to pick and choose 3-4 components using their specific icon, which were then moved by the facilitators to the next room.   
    \item \textbf{DIY Station (Floor 3 West)---Generation (30 mins):} In a breakout room, this section of the workshop is designed for a participant to conceptualize and generate an ethical action plan using: the collected artifacts from the \textit{Shop} (previous stage), one method of their choice that they currently use in everyday work, and some other provided building blocks such as potential actors and relational verbs. 
    The ``Problem Card'' from the \textit{Prep Room} (second stage) is also presented for the participant's reference as they create the action plan to address the problem they identified initially. A feedback loop is introduced in the middle of this stage for the participants to quickly share what they are in process of building with a fellow participant to allow them to be aware of what their co-participants are building. 
    \item \textbf{Test Drive Room (Floor 3 East)---Evaluation (30 mins):} In a different breakout room, this section of the workshop is designed for a participant to evaluate and iterate upon the action plan they created in the previous stage. To evaluate, the participant is first provided with material to create a new ``Context Card'' that will provide a different perspective than the original ``Problem Card,'' for example, by identifying a new setting (e.g., team meeting, heads-down time, client meeting) to evaluate how the action plan might function in this new context. The participants are provided with colored Post-Its to mark aspects of their action that ``passed,'' ``failed,'' and ``could be improved.'' 
    \item \textbf{Gallery (Floor 4)---Final Presentation:} This section of the workshop marks the completion of the workshop where each participant is given time to present the action plan they created during the workshop. Each participant describes their Problem Card, Action Plan, iterations in the Action Plan, and learnings from the workshop. 
\end{itemize}

We leveraged the design outcomes and process from Case A and Case B to construct and situate a range of \textit{facets of designing for co-design} that could be referred by co-design researchers and designers as a form of intermediate-level knowledge, described in detail in the next section. 

\section{FACETS OF DESIGNING FOR CO-DESIGN} \label{sec:facets}
\begin{figure}[ht]
    \centering
    \includegraphics[width=\textwidth]{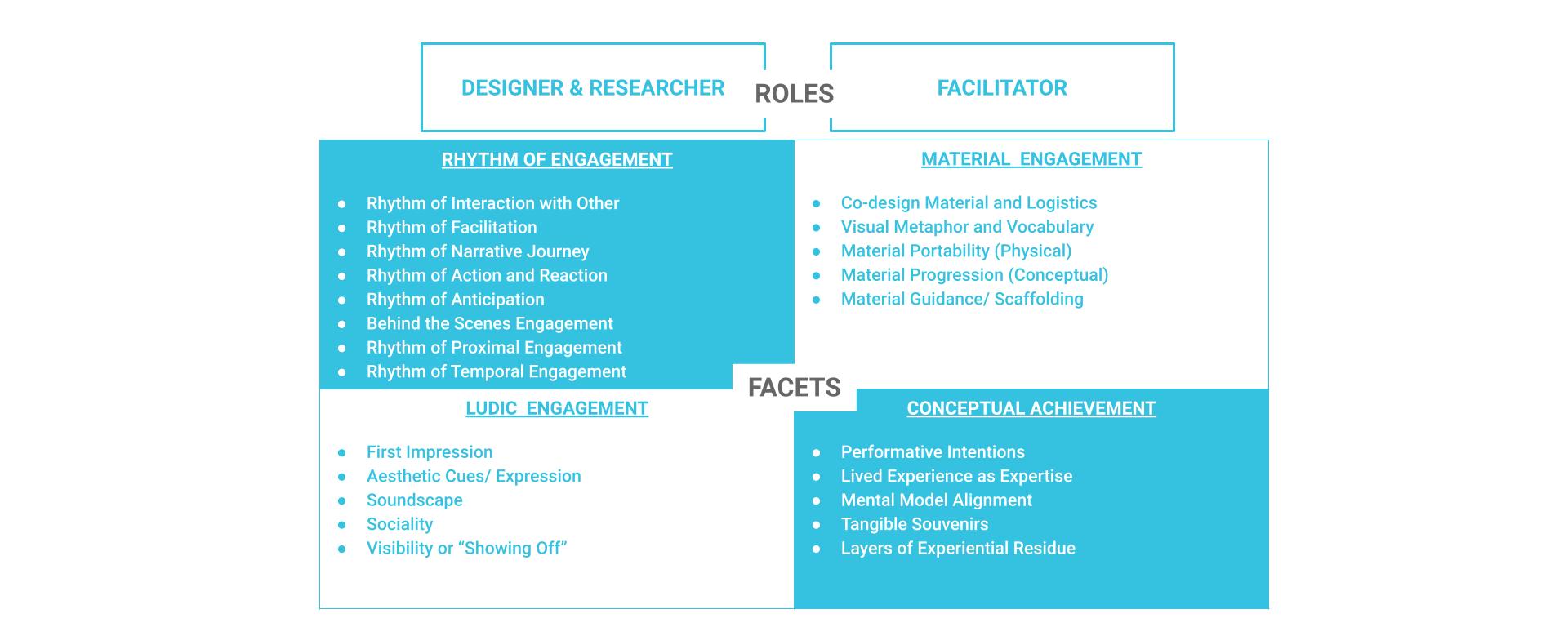}
    \caption{Facets and respective sub-facets that Frame Co-Design as a Designerly Practice}
    \label{fig:facets}
    \Description[Schema]{Figure consisting of a schema listing all the four facets, sub-facets under each facet, and three roles involved in designing for co-design. The schema has four big rectangles placed in a 2X2 matrix. Going clock-wise, the rectangles list the sub-facets in text for the facets- Rhythm of Engagement, Material Engagement, Conceptual Achievement, and Ludic Engagement. At the top of this 2X2 matrix arrangement, are two smaller rectangles (with text ``Designer \& Researcher'' and ``Facilitator'') covering the roles involved.}
\end{figure}

Through the process of designing the co-design experiences described in ~\ref{sec:casestudydescriptions}, we constructed four primary \textit{facets} that form a novel set of intermediate-level knowledge that can support co-design research and design. These facets, shown in Figure~\ref{fig:facets}, include knowledge along the dimensions of: 1) \textit{Rhythm of Engagement} over space and time; 2) \textit{Material Engagement} through interaction in a co-design space; 3) \textit{Ludic Engagement} to implement playfulness; 4) \textit{Conceptual Achievement} to scaffold and direct outcomes; and 5) \textit{Human-as-Facilitator} to highlight the engagement of humans.

\begin{table*}[hb]
\centering
\caption{Description of Facets that Frame Co-Design as a Designerly Practice (listed in Figure \ref{fig:facets})}
\label{Codebook:Facets}
\Description[Table with two columns]{Table with two columns. First, ``Facet'' that identifies an aspect of designing for co-design, ``Description: Facet of Co-design that caters to \ldots''. The rows of the table consists of four rows for each facet described in detail in the paper and its description. First row caters to ``Rhythm of Engagement.'' Second row describes the facet ``Material Engagement.'' Third rows describes ``Ludic Engagement.'' Fourth row describes ``Conceptual Achievement.''}
\begin{tabularx}{\textwidth}{p{.25\textwidth}p{.7\textwidth}}
\toprule
\textbf{Facet} & \textbf{Description: Facet of Co-design that caters to\ldots}\\ \midrule
\textbf{Rhythm of Engagement} & designing of the orchestration of temporal, spatial, and conceptual composition of various activities, types of engagement, and/or interactions throughout the workshop time. \\
\textbf{Material Engagement} & designing for tangible material qualities, interaction, and movement. \\
\textbf{Ludic Engagement} & designing aesthetic experiential qualities of a workshop including elements of fun, entertainment, sociality, networking, and visibility to one's co-designed material.\\
\textbf{Conceptual Achievement} & designing both tangible and experiential sense of knowledge production, sharing, and evaluation. \\
\bottomrule
\end{tabularx}
\end{table*}

Each facet dimension includes a variety of sub-facets. Each sub-facet is illustrated through our design process from Cases A and B, including relevant reflection, derivation, and/or illustration of the sub-facet. We provide a definition of each sub-facet in \textbf{bold} text and conclude with some questions you might want to ask yourselves as researchers or designers of co-design experiences in \textit{italicized} text.

We also rely upon a glossary of common terms relating to the design and experience of co-design environments, which we include for clarity below:

\begin{itemize}
    \item \textit{Co-design space:} Sanders and Westerlund \cite{Sanders2011-wm} define co-design spaces across three layers: 1) experienced \textit{physical} space; 2) a \textit{metaphorical} space to encourage the participants to share their ideas and experiences; and 3) resultant co-designed situations (ideas or future scenarios and fears). Here, co-design spaces include the physical and/or digital environment which might include Zoom breakout rooms, main rooms, physical spaces, and the entire workshop as a context in which the participants as people can interact and co-create.
    \item \textit{Co-design material:} Co-design frames a user's involvement through facilitation by the designer or researcher ``\textit{through sharing the practical tools that can be used to enable participation, collaboration and creative thinking}'' \cite{Blomkamp2018-iv}. Here, co-design material that is designed for participants includes probes, toolkits, prompts, scaffolds, or other elements that are previously designed, provided, and/or interacted with by facilitators and participants. This also includes material that is specific to facilitators such as Session scripts, private communicative channels, and music or other audio prompts. We discuss this in much detail through sub-facets under Material Engagement in Section \ref{sec:materialengagement}.
    \item \textit{Designers:} Co-design researchers who intentionally design co-design space(s) and material(s). We discuss this in much detail as we present the ``hats'' during co-design in Section~\ref{sec:hats}.
    \item \textit{Cases:} Moving forward, we refer to these as Case A and Case B to point towards specific examples of design processes and decisions that led us to characterize a specific facet. 
\end{itemize}


\subsection{Rhythm of Engagement}
Rhythm of Engagement is a facet of co-design that caters to designing of the orchestration of temporal, spatial, and conceptual composition of various activities, types of engagement, and/or interactions throughout the workshop time. It includes how the workshop session (120-180 minutes) is arranged such as interactions among workshop participants, facilitation logistics, various actions to be taken by the participants in the workshop, and timing for each of these sub-activities that forms a holistic experience with the materials, knowledge, people, and themselves.

\begin{table*}[hb]
\centering
\caption{Facet: Rhythm of Engagement}
\label{Codebook:Rhythm}
\Description[Table with two columns]{Table with two columns to present a list of sub-facets under the facet ``Rhythm of Engagement.'' First column, ``Sub-facet'' lists different items under the main facets. Second column, ``Description'' defines each sub-facet. The rows of the table consists of eight rows for each sub-facet described in detail in the paper and its description.}
\begin{tabularx}{\textwidth}{p{.25\textwidth}p{.7\textwidth}}
\toprule
\textbf{Sub-Facet} & \textbf{Description}\\ \midrule
\textbf{Rhythm of Interaction with Others} & \ldots includes combinations of heads-down time for one participant, paired-up activities for two or three participants (in a sub-group or breakout room), or open group interactions with all the participants in the workshop. \\
\textbf{Rhythm of Facilitation} & \ldots includes the degrees of involvement of workshop facilitators within the co-design space, material, and participant's activities. \\
\textbf{Rhythm of Narrative Journey} & \ldots includes a characterization of the overall experience designed for participants, including elements of the narrative arc which are either slowly revealed over time or otherwise sequenced.\\
\textbf{Rhythm of Action and Reaction} & \ldots includes combinations of postures towards other participants or the material provided/created throughout the workshop including self-reflecting on one's own creation, reflecting on other's work, guiding fellow participants through an activity/reflection, observing others' creation, providing feedback on other's work, and evaluating or re-designing one's own work after feedback, reflection, and/or observation.  \\
\textbf{Rhythm of Anticipation} & \ldots includes expectations of the workshop experiences and potential participant needs that might conflict with workshop plans,including ways the workshop might be modified.  \\
\textbf{Behind the Scenes Engagement} & \ldots includes facilitator interaction over time through one or more communication channels to ensure all the facilitators are aligned with the planned and emergent rhythms of engagement, address contingencies in sub-groups, and engage back-and-forth between the sub-groups and main co-design space.  \\
\textbf{Rhythm of Proximal Engagement}  & \ldots includes planned opportunities for participants to move towards and away from the co-design material throughout the workshop. \\
\textbf{Rhythm of Temporal Engagement} & \ldots includes the arrangement of interactions and actions over time.  \\

\bottomrule
\end{tabularx}
\end{table*}

\subsubsection{Rhythm of Interaction with Others}
In Case B, we sought to carefully craft the kinds of interactions a workshop participant has with other participants in the workshop to balance and proactively manage cognitive, intellectual, and social energy. We planned when the participants had ``downtime'' to think alone, including individually creating their problem cards, shopping the method components, creating their action plan, and iterating on their action plans. In parallel, to align with the participatory elements of a co-design session, we sought to include more socially-interactive elements such as: 1) interactive sharing in the Lobby Room for participants to interact with each other's problem cards, 2) shopping at the same time, while individually collecting method components, creating the ``buzz'' of collective shopping experience, and 3) break-time discussions to encourage discussion of participants' progress and ideas around their action plan that would support further downtime. In Case A, this sub-facet was not considered as it included only one participant with one facilitator and was designed to be more conversational which had higher engagements with the facilitator than in Case B. 
\textbf{Rhythm of interactions with other participants includes combinations of heads-down time for one participant, paired-up activities for two or three participants (in a sub-group or breakout room), or open group interactions with all the participants in the workshop.} As designers, this sub-facet encouraged us to consider:\textit{ When to position an activity (timing); what we wanted the participants to do and in what combination (interaction); and where in the co-design space the interaction should occur (in separate group/breakout room or overall group/main room)?}

\subsubsection{Rhythm of Facilitation}
In Case A, the interaction between the workshop participant and the facilitator was one-on-one and more conversational which required the constant involvement of the facilitator. By contrast, in Case B, the interaction between the workshop participants and the facilitator was often more distant and punctuated, including moments where the facilitator provided instructions at the beginning of the workshop, to support conversation among participants in sharing periods, and when there was a question from a participant. The interactive rhythm of these points of facilitator involvement were used to highlight participants' engagement and discussion with each other.
\textbf{Rhythm of facilitation includes the degrees of involvement of workshop facilitators within the co-design space, material, and participant's activities.} This difference in the roles and engagement of the facilitator, experienced over time, made us think about: \textit{When do we want the facilitators to guide the participants (or not)? How much do we want the facilitators to be involved with the participant's interaction with others and the material?}

\subsubsection{Rhythm of Narrative Journey}
At the workshop level, in Case A, participants had different combinations and sequences of action as they physically mapped (as in Activity A), narrated stories (as in Activity B), or annotated a method (Activity C), thereby reflecting upon ethical dimensions of their practice through different forms of interaction with the co-design material. In Case B, we designed for a defined flow of action for the participants to: 1) familiarize themselves with their practice, 2) identify a problem they want to build an action plan for, 3) collect the components, and then, 4) generate an action plan, to further 5) test and evaluate for their everyday practice. In Case B, the flow of action was slowly built up from the beginning of the workshop to the end, starting from primarily thinking and evaluating to doing hands-on work and celebrating final outcomes; in contrast, the narrative arc of Case A was focused entirely on evaluative and evocative work to describe the ethical complexity of the participant's work practices based on different participants being introduced to different sequences of activities. These two different cases describe differing means of constructing a narrative arc to build up and shape the role and intended destination for the participant.

\textbf{Rhythm of Narrative Journey for the participants includes a characterization of the overall experience designed for participants, including elements of the narrative arc which are either slowly revealed over time or otherwise sequenced.} As designers, we had to negotiate the kind of journey we wanted the participant to have, asking: \textit{How do we want to build the experience from step-to-step for the participant? What are the desired relationships of various interactive moments to the narrative whole? What kinds of journeys are more appropriate with regard to the workshop goals?}

\subsubsection{Rhythm of Action and Reaction}
In Case A, participants interacted with the co-design material and simultaneously reflected and reacted back to the material through conversation with the facilitator. In Case B, due to multiple participants engaging with the co-design material in different spaces, we intentionally designed spaces and social settings (e.g., the Lobby, DIY room evaluation, Gallery Room) where participants had a chance to not only engage with their own created material but also give feedback, reflect on, and discuss other participants' artifacts. In these cases we considered the desired rhythm of action (initial design) and reaction (co-design/iteration/reflection) with the goal of avoiding ongoing saturation in either of these modes. In Case A, this rhythm of action and reaction was designed to happen simultaneously, whereas, in Case B, the rhythm was designed to occur through multiple chains of action/reaction across the three hour workshop. 
\textbf{Rhythm of action and reaction includes combinations of postures towards other participants or the material provided/created throughout the workshop including self-reflecting on one's own creation, reflecting on other's work, guiding fellow participants through an activity/reflection, observing others' creation, providing feedback on other's work, and evaluating or re-designing one's own work after feedback, reflection, and/or observation.} As designers, this made us think about:\textit{ What are the different ways we can provide time for the participant to engage with their own and others' work throughout the workshop? How can we use a shift in mode to avoid cognitive or social fatigue?} 

\subsubsection{Rhythm of Anticipation}
In Case A, we initially did not prepare for the participant to need to hold a newborn baby during the session---causing the facilitator to become the hands of the participant, and shifting a lot of expectations, roles, and work. In this extreme case, the shift in responsibilities also led to dividing the workshop session into two different days considering the sudden needs of the participant. In Case B, we anticipated that the participants may want to go back and change their method components collection (where they can pick 3-4 components) while designing their action plan; although it would have been more straightforward to design this portion of the workshop more rigidly, we decided to allow flexibility to the participants, but as a ``back-rhythm'' in our session script that we had to consider as a part of the overall rhythm, disrupting---but potentially enriching---the overall narrative rhythm. Especially in Case B, given the presence of multiple sub-groups and spaces in the workshop, we also anticipated misalignment across subgroups in terms of time and expectations, resulting in the addition of buffer time in our session script (anticipation in Rhythm of Temporal Engagement), backchannels among the facilitators (anticipation in Behind the Scenes Engagement), and re-grouping in case of a no-show from the participants (anticipation in Rhythm of Interaction with Others).
\textbf{Rhythm of Anticipation includes expectations of the workshop experiences and potential participant needs that might conflict with workshop plans,including ways the workshop might be modified.} As designers, we have to be prepared with additional set of instructions (in a way that won't lead the participants); re-routing the rhythm to ensure alignment among the participants in relation to the overall goal of the workshop; and a set of potential reactions from the facilitators to make the participants feel ownership in the contingency plan. We asked ourselves: \textit{What potential (unplanned) participant reactions might occur at each stage? Where will we need alternatives based on participants' potential reactions? What preparation will the facilitators need for the contingency pathways?}

\begin{figure}
    \centering
    \includegraphics[width=\textwidth]{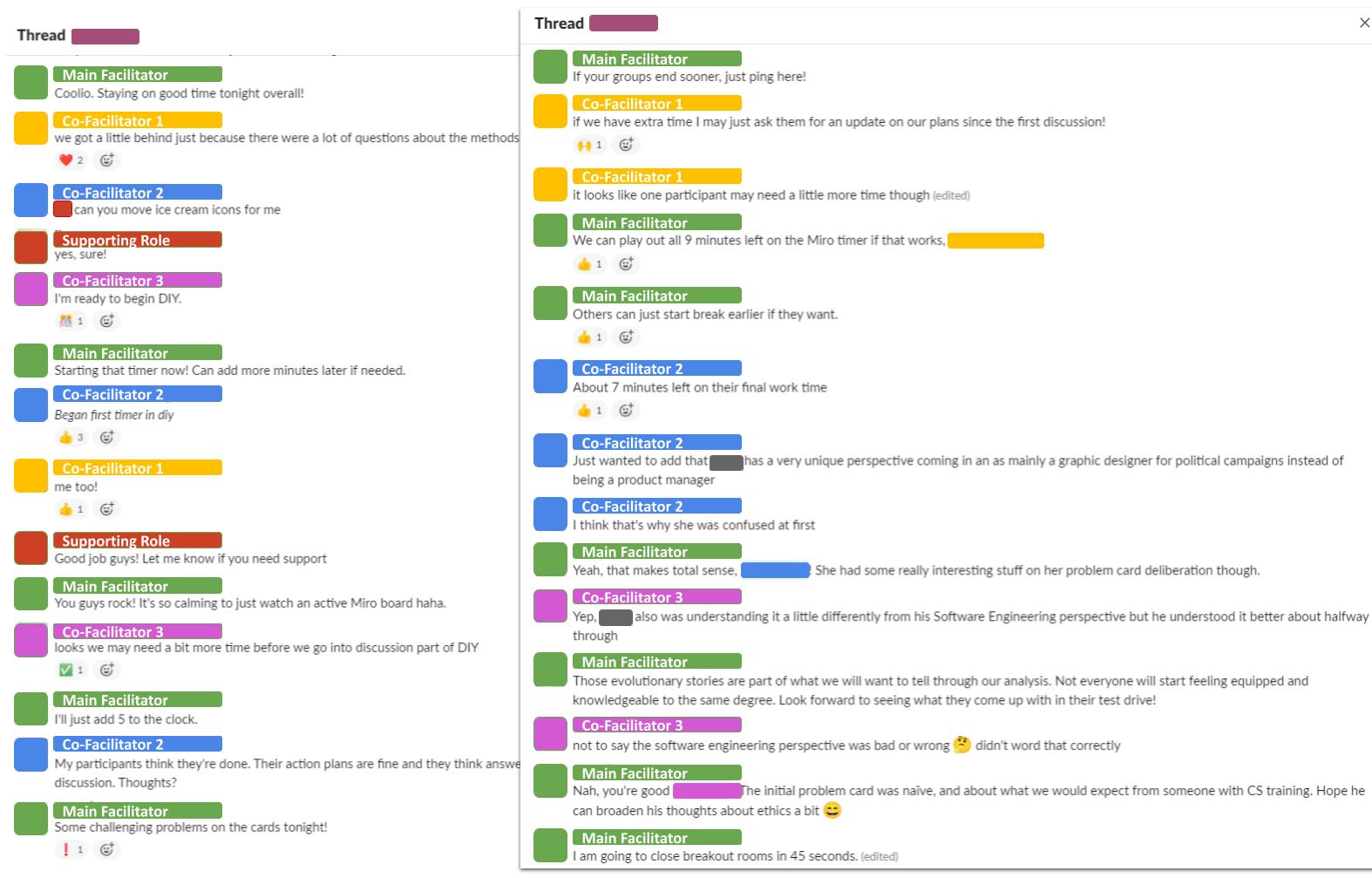}
    \caption{Snippet of a Slack Thread during a Case B Workshop to illustrate Behind-the-Scenes Engagement; This conversation illustrates the co-facilitators from different rooms coordinating during a transition time in the Session Script.}
    \label{fig:btsslack}
    \Description[Figure of two Screenshots]{Figure having two screenshots of a Slack thread of a conversation between Main facilitators and three co-facilitators. The names and images of the actual facilitators are covered using colored blocks and represented as Co-facilitator 1, 2, and 3.}
\end{figure}

\subsubsection{Behind the Scenes Engagement}
In Case B, due to the use of breakouts and a main room, we had a main facilitator for the entire workshop and 3-4 co-facilitators for subgroups. This required behind-the-scenes coordination to align various subgroups in the entire co-design space. We set up a Slack thread (as shown in Figure \ref{fig:btsslack}) to make sure all facilitators were connected ``live'' as we worked through the session script in parallel. The backchannel helped us to form contingency plans in real time, such as extension of time for particular rooms, conversation among facilitators to make sure all participants were given similar information, and discussion of participant questions to align responses. In Case A, there was minimal need for this sub-facet, since there was only one facilitator, which reduced the complexity of coordinating across multiple facilitators. In this case, behind the scenes activities were more logistically-focused, including recording, timer set-up, screen-sharing, script references, and taking notes.
\textbf{Behind-the-Scenes Engagement includes facilitator interaction over time through one or more communication channels to ensure all the facilitators are aligned with the planned and emergent rhythms of engagement, address contingencies in sub-groups, and engage back-and-forth between the sub-groups and main co-design space.} In a physical space, engagement in ``behind the scenes'' might manifest in a 5-minute side conversation while facilitators huddle to plan for contingencies or alignment. As designers, we had the main facilitator set up a Slack thread and often asked questions in real-time such as: \textit{Do we need more time? Are we ready to move on to the next part in our Session Script? Do we need any assistance for facilitation in any sub-group? When should we provide reminders of starting and stopping the Zoom recording? What ``juicy quotes'' or notes of observation should we document in real time?} 

\subsubsection{Rhythm of Proximal Engagement}
In both Cases A and B, given the digital medium of engagement in co-design, we had to plan the proximal engagement of the participants to include robust amounts of off-screen time (such as breaks), off-activity times (such as reflection and storytelling times which did not include looking at the screen but rather focused attention on a verbal conversation), off-interaction times (such as individual down-time away from the group or facilitators). Additionally, by using the zooming affordance in Miro, we were able to create more or less distance from specific activities, and in Case B, using this zooming to reinforce progression through the workshop (zooming into one room at a time) and relationships among activities. Similarly in physical space, proximal engagement might include allowing for a ``scene change'' or for participants to mode-shift after intense periods of work---marked by distance in physical space rather than in digital space.
\textbf{Rhythm of Proximal Engagement includes planned opportunities for participants to move towards and away from the co-design material throughout the workshop.} As designers, we considered: \textit{When is the right time for a break? Where are the potential fatigue moments in the workshop that we need to mitigate with a different rhythm of activities or a quick break? What points in the workshop timeline should we prioritize screen-time breaks? How should we plan to manage energy levels of rest vs. intellectual drive? }

\subsubsection{Rhythm of Temporal Engagement}
In Case A, we designed two different activities that the participant engaged in over a 90--120 minute period, using a sequence to relate or ``tie together'' the activities based on their temporal proximity in a concluding reflection. In Case B, we designed a linear progression of narrative where the participant moved through stages of problem creation, conceptualization, iteration, and presentation of their action plans over a 120--180 minute period. Given the long periods of time engagement, it required us to plan the temporal engagement of each smaller activity, with specific time intervals, to create a balance of energy, reflection, discussion, and participation for the participant---considering how components of the workshop related to elements that were temporally ``next'' to each other, and how all of the elements contributed to an overarching temporal rhythm.  
\textbf{Rhythm of Temporal Engagement includes the arrangement of interactions and actions over time.} As designers, this encouraged us to plan: \textit{How much time to provide for different actions? How much time to allot for different interactions? When do we want participants to notice ``serendipitous'' connections among activities?}

\subsection{Material Engagement}\label{sec:materialengagement}
Material Engagement is a facet of co-design that informs the design of tangible material qualities, interactions, and movement, encompassing: 1) tangible materials such as logistics documents and the visual language of the materials; 2) conceptual and physical movement of the material during the workshop in relation to participant and facilitator interactions; and 3) sense-making through material scaffolds provided to the participants to aid them in achieving the workshop goals.

\begin{table*}[hb]
\centering
\caption{Facet: Material Engagement}
\label{Codebook:material}
\Description[Table with two columns]{Table with two columns to present a list of sub-facets under the facet ``Material Engagement.'' First column, ``Sub-facet'' lists different items under the main facets. Second column, ``Description'' defines each sub-facet. The rows of the table consists of five rows for each sub-facet described in detail in the paper and its description.}
\begin{tabularx}{\textwidth}{p{.25\textwidth}p{.7\textwidth}}
\toprule
\textbf{Sub-Facet} & \textbf{Description}\\ \midrule
\textbf{Co-design Material and Logistics} & \ldots includes any tangible information used in the co-design space including session scripts, consent forms, physical/digital spaces, audio and video recordings, sheets of instructions, prompts, toolkits and probes, workshop take-aways and outputs. \\

\textbf{Visual Metaphor and Vocabulary} & \ldots includes approaches to visually and conceptually organize information and provide a shared vocabulary for interaction. 
\\

\textbf{Material Portability} & \ldots includes the physical and spatial movement of co-design materials, allowing participants to access relevant information to inform a task, or to relate present activities to previous artifacts for further manipulation or refinement. \\

\textbf{Material Progression} & \ldots includes the interactive relationship between material affordances and scaffolds to inform a sense of material progression during the workshop.  \\

\textbf{Material Guidance/ Scaffolding} & \ldots includes visual and textual affordances to support participants in performing intended actions at each stage of the workshop. \\

\bottomrule
\end{tabularx}
\end{table*}

\subsubsection{Co-design Material and Logistics}
\begin{figure}
    \centering
    \includegraphics[width=0.7\textwidth]{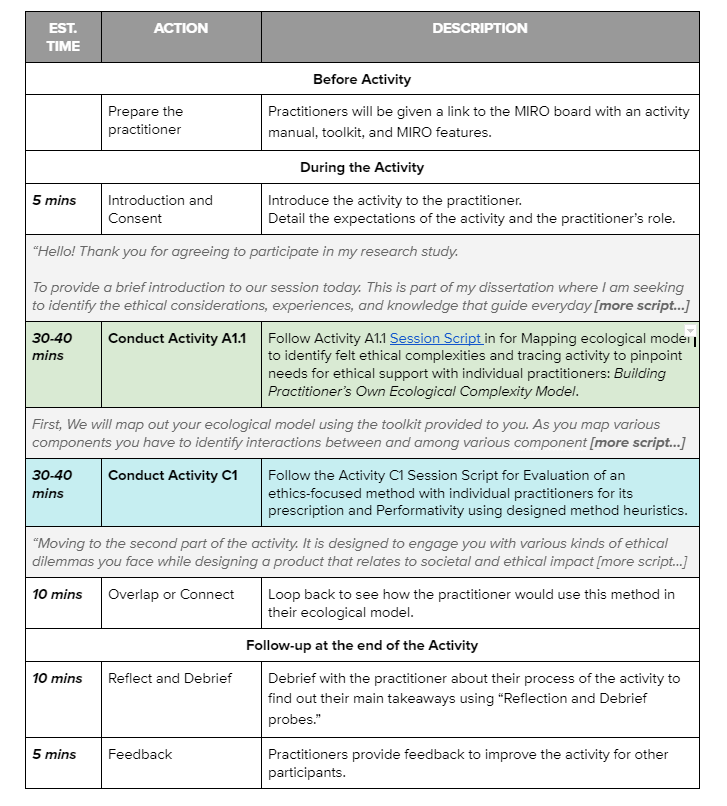}
    \caption{Example of a Session Script of a Workshop}
    \label{fig:sessionscript}
    \Description[Figure of Snippet of a document]{Figure consisting of a snippet of a Session Script example. The snippet has a representation of a table with three columns--- Estimated time, Action, and Description. The text in each row corresponds to the chronological order of how a co-creation session would progress as a plan presented in a Session Script.}
\end{figure}

In Cases A and B, we designed a range of physical and digital materials to prepare for the session (as facilitators) and to scaffold co-design engagement (by participants), such as: 1) session scripts and narrative scripts which included sample timing and role details (as in Figure \ref{fig:sessionscript}); 2) space setup in the digital environment including Miro boards, Zoom rooms, Google Docs, and Slack workspaces; and 3) co-design materials, including sheets of instructions, prompts, and probes and toolkits for participants to leverage as they engaged in a range of activities. In both cases, we considered these materials as touchpoints that informed the overall cohesion of the workshop, encouraged consistency across sessions, and enabled us to go ``off script'' where needed to encourage flexibility of facilitation and participation roles. 
\textbf{Co-design Material and Logistics includes any tangible information used in the co-design space including session scripts, consent forms, physical/digital spaces, audio and video recordings, sheets of instructions, prompts, toolkits and probes, workshop take-aways and outputs.} The creation of logistics- oriented material  ensured us to have uniformity across different workshops and among co-facilitators. As designers, we sought to create materials to support a range of different roles: \textit{As a researcher, what materials do we need to have in place to make the workshop happen? As designers, what co-design materials do we need to provide the participants to achieve the workshop goals? As facilitators, what logistics-related material do we need to prepare to ensure uniformity and smooth workshop facilitation?}

\subsubsection{Visual Metaphor and Vocabulary}
In Case B, we designed a co-design environment to guide the participant through the creation of their own action plan to support professional ethics. To make this overall narrative fun, interactive, and meaningful, we ideated using different metaphors to frame elements of the workshop, with inspiration from the grocery cart/shopping experience, art/craft studio, shopping mall, and museum. Through these metaphors, we were able to identify coherent step-by-step progression to aid the participants in achieving their desired task. Inspired by these metaphors, we finalized our materials using several key metaphors. First, we used the metaphor of crafting a prototype in the DIY room, highlighting the metaphorical objective of ``assembling a toolbox.'' Second, we sketched a ``building'' with different ``rooms'' (Figure \ref{fig:CaseB}) for each step with elements that were to be carried forward from one room to another. We named each room to give a sense of the potential actions to be performed at that stage of the co-design workshop such as a ``Shop'' for collecting artifacts required for ``DIY room'' where participants had to create their own action plan using the collected artifacts. Based on feedback from the first two pilot sessions, we changed the language of the workshop objective from ``creating a method'' (which was difficult for participants to understand), to ``assembling your toolkit,'' and finally ``creating your action plan.'' 
In Case A, we iteratively refined specific vocabulary used around ethics, given the range of professional roles intended to engage with the toolkits, such as ``ethical dilemmas'' (in Activity B) and ``ecological complexity informing ethics'' (in Activity A). 
\textbf{Visual Metaphor and Vocabulary includes approaches to visually and conceptually organize information and provide a shared vocabulary for interaction.} As designers, we constantly evaluated: \textit{How are the participants going to make sense of the co-design material provided to them? What mental models might participants use to connect multiple elements of the co-design experience? How can we make the material more accessible to a range of participant types without excessively didactic facilitation?}

\subsubsection{Material Portability}
In Case B, participant began the process by creating a ``Problem Card'' in the Prep Room (first step) which was then moved to the Lobby (second step). This material later formed a design frame as they created their action plan in the DIY Room, tested their action plan in the Test Drive Room, and presented in the Gallery. The need to move this material caused our group of designers to consider the role of visual placeholders (represented in dotted lines in Figure \ref{fig:CaseB}) in the material design to incorporate the moved material directly within the visual theme. We designed the portability of material in real time within the different co-design spaces as well, allowing for participants to take the co-design material ``home'' with them after the conclusion of the workshop (further discussed in Conceptual Achievement---Tangible Souvenirs). In Case A, we prepared for portability, allowing participants to overlap their filtered ethical dilemmas (in Activity B) over a previously mapped ecological complexity (in Activity A). The portability requirements allowed us to consider mechanisms to support the relative scale of the material, affordances that indicated portability to participants and facilitators, and visual placeholders in our co-design materials. 
\textbf{Material Portability includes the physical and spatial movement of co-design materials, allowing participants to access relevant information to inform a task, or to relate present activities to previous artifacts for further manipulation or refinement.} As designers, we considered: \textit{What visual placeholders in our material design could incorporate the moved material within the visual metaphor/theme? How can the portability of materials allow past interactions to inform later interactions in the workshop? Who is responsible for conducting the movement of materials during the workshop?} 

\subsubsection{Material Progression}
In parallel with the physical material portability, material transposition facilitated conceptual movement across the narrative in physical form. 
In Case A, we encouraged the participants to connect ethical dilemmas (in Activity B) they faced to various components and connections in their ecological map (in Activity A)---a conceptual mapping which was not visually articulated, but made prominent through the relationship of frames on the Miro board. In Case B, we asked participants to create a new ``Context Card'' as a frame that described the purpose and intended goal of their toolkit, building on a previously designed ``Problem Card'' constructed early in the workshop, which had to be carried forward from Prep room (in first floor) to Test Drive Room (in third floor). In this latter example, the shift in naming of the card and the desire to create contrast with previous efforts to frame design activity provided conceptual movement that was strengthened by the material progression. 
\textbf{Material Progression includes the interactive relationship between material affordances and scaffolds to inform a sense of material progression during the workshop.} As designers, we considered ways to provide links across the workshop to consider: \textit{How does one activity or step lead to the next or the other conceptually? How can we convey that movement through the material, both visually and conceptually?}

\subsubsection{Material Guidance/ Scaffolding}
In Cases A and B, we designed affordances into the co-design material to connect the overall narrative arc and interactions the participant needed to perform. For instance, we used flexible and moveable Post-Its for \textit{annotating} (Case A---Activity C), empty stacks of post-its (Case B---Ground Floor) to indicate \textit{multiple desired responses}, elicitation dilemma cards and postcards (Case A---Activity B) for \textit{answering} specific questions with multiple responses, and flow chart components to support \textit{freeform creation and mapping} (Case A---Activity A and Case B---DIY Room). 
In Case B, we created scaffolding by naming different rooms (sub-spaces) to frame their interaction in that space through a particular intention to be achieved by the participant. In Case A, the one-on-one interaction between the facilitator and the participant reduced the efforts to create guidance around the scaffolds. 
\textbf{Material Guidance/ Scaffolding includes visual and textual affordances to support participants in performing intended actions at each stage of the workshop.} The scaffolding is embedded into the probes, toolkits, and/or activities to allow the participants co-create their knowledge through the co-design material. As designers, we evaluated the efficacy of the affordances, asking: \textit{What are the intended actions we would like the participants to perform through the provided toolkits? What kinds of guidance should we provide them to interact with the co-design material?}


\subsection{Ludic Engagement}
Ludic Engagement is a facet of co-design that caters to designing aesthetic experiential qualities of a workshop including elements of fun, entertainment, sociality, networking, and visibility to one's co-designed material.

\begin{table*}[ht]
\centering
\caption{Facet: Ludic Engagement}
\label{Codebook:ludic}
\Description[Table with two columns]{Table with two columns to present a list of sub-facets under the facet ``Ludic Engagement.'' First column, ``Sub-facet'' lists different items under the main facets. Second column, ``Description'' defines each sub-facet. The rows of the table consists of five rows for each sub-facet described in detail in the paper and its description.}
\begin{tabularx}{\textwidth}{p{.25\textwidth}p{.7\textwidth}}
\toprule
\textbf{Sub-Facet} & \textbf{Description}\\ \midrule
\textbf{First Impression} & \ldots focuses on elements of onboarding and other means of ``setting the tone'' of the workshop, including initial felt impressions and how those impressions shape potential interactions for the group of the participants.  \\

\textbf{Aesthetic Cues/ Expression} & \ldots focuses on the creation of ludically-focused aesthetic elements based on visual metaphors or material shared in the co-design space with the participants.\\

\textbf{Soundscape} & \ldots includes the selection of auditory cues to set the mood and provide a focused rhythm for the participants, supporting individual, collaborative, and design-focused work periods. \\ 

\textbf{Sociality} & \ldots focuses on creating opportunities for networking and/or structured informal interaction among the participants to share information or personal stories.\\

\textbf{Visibility or ``Showing Off''} & \ldots includes means for participants to present or otherwise share their work, thereby creating opportunities for reflection, engagement, and collaboration using the co-designed material as a platform for engagement within and/or across a group of participants. \\ 


\bottomrule
\end{tabularx}
\end{table*}

\subsubsection{First Impression}
In both Cases A and B, we scripted the onboarding to the workshop to ensure that the participants felt as if they belonged in the co-design space, organically ``setting the stage'' to frame their expectations of the workshop. 
In Case A, the onboarding experience was conducted one-on-one with the facilitator, removing the complexity of being a part of an unknown group of people. The introductory script framed the interaction as follows: ``\textit{this is your space and you are welcome to share what your practice looks like. You are sharing your expertise with me},'' which put the participants at ease and provided them with a frame to participate. In Case B, given the multiple participants, we also had to make sure participants felt they were part of the group and comfortable as they started working with each other in a breakout room. We ideated on a range of ice-breaking activities and considered multiple ways to introduce the visual ``house'' metaphor, with the goal of building curiosity in the participants before bringing them fully on board. The initial visual metaphor of the ``house'' may have also made the participants feel overwhelmed---with their goal to to metaphorically climb ``four floors'' over a three hour period---so we had to carefully craft our introductory script to be supportive of all these factors. 
\textbf{First Impression focuses on elements of onboarding and other means of ``setting the tone'' of the workshop, including initial felt impressions and how those impressions shape potential interactions for the group of the participants.} These introductory steps to ease participants into the workshop can take the form of simple tutorials of co-design material, fun introductions from both facilitators and participants, friendly ice-breaking prompts, etc. As designers, we considered: \textit{How can we create a friendly start to the session? How can we not overly force casual interactions yet encourage participants to feel comfortable among each other? How can we balance feelings of being overwhelmed with feelings of possibility and opportunity? 
What should we do to make participants feel like experts of their own experience and lay the groundwork for exciting co-design ``vibes''?}

\subsubsection{Aesthetic Cues/Expression}
In Case B, we received almost-immediate, positive responses from participants who felt very excited by their ability to zoom in and out of the ``house,'' searching across the ``floors'' for the next activity. Even across an intensive three hour workshop, participants were very enthusiastic as they were introduced to each room and as we took them to the ``next floor up the stairs.'' This ludically-focused visually metaphor connects back to our support for \textit{Material Engagement}, through which we sought to provide an aesthetic experience for the participants. Beyond framing visual metaphors, we also used other small hooks for ludic engagement in Case B, including the inclusion of small stick figures on some of the virtual stairwells, and encouragement for participants to draw ``outside of the lines'' to make the space their own.
\textbf{Aesthetic Cues/Experience focuses on the creation of ludically-focused aesthetic elements based on visual metaphors or material shared in the co-design space with the participants.} As designers, we designed for an aesthetic experience by asking: \textit{Are we providing a visually exciting experience? How can we sustain participants' curiosity for longer periods of time? How can we convey the purpose of a stage in the workshop in a fun way to allow the participants to easily connect to the intent? How can we encourage participants to play and ``break the rules''?}

\subsubsection{Soundscape}
In Case B, we used sound to signal the progression through multiple stages in the workshop, including instances where the participants had downtime in breakout rooms, times where they intended to move around to view others' problem cards (in the Lobby) or collect method components (in the Shop), or focused individual co-design time (in the DIY Room). We intentionally incorporated a range of music to support these periods, including a combination of relaxing, ambient music for focus time, and fun/ energizing music for breaks and moments of collaboration. 
We also included the ``Timer'' sounds on the Miro board to give participants a sense of time, allowing all participants across different sub-spaces to align in terms of the temporal rhythm. 
\textbf{Soundscape includes the selection of auditory cues to set the mood and provide a focused rhythm for the participants, supporting individual, collaborative, and design-focused work periods.} As facilitators, we considered: \textit{Should the music demand attention in the foreground or shift towards the background based on the interactions in the rhythm of engagement? How can music be used to establish interactional norms and build a sense of expectation? How can soundscapes be used to temporally align activities and build a shared sense of interaction?}

\subsubsection{Sociality}
In Case B, we incorporated informal periods for leisure as a way to give participants a break from focused co-design work. In a physical set-up, this would have been easier due to the physical presence of people in a room, but in a digital environment, we had to be intentional and explicitly introduce and support opportunities for sociality. For instance, prompts for sociality could include dramatic and complete breaks from the session---indicated by music or through an invitation to step away from a Miro frame where work was taking place---or more informally, by encouraging participants to chat amongst themselves before they transitioned to the next phase of the workshop. While we intentionally built opportunities for sociality, we also recognized that digital interactions are draining, and so while we designed an environment for the participants to network and know more about each other, we wanted them to do this in relation to their own comfort and choice, recognizing ``no sociality'' as a legitimate outcome for some participants. In Case A, we did not directly consider participant-participant social elements as this workshop was a one-on-one interaction with the facilitator. These social elements were more embedded and naturally occurred in the conversation between the participant and the facilitator.  
\textbf{Sociality focuses on creating opportunities for networking and/or structured informal interaction among the participants to share information or personal stories.} This comfort across the participants is later projected in their interactions in their sub-groups. As designers, we considered: \textit{What are some avenues we can provide for the participants to interact beyond the co-design space and activities? How can we allow participants to ``hang-out'' with each other in break times (given the digital medium) if they choose to do so and still balance the Rhythm of Proximal Engagement? How can the digital medium of the workshop constrain or enable sociality? How do we protect individuals that are exhausted from feeling as if they have to be social?}

\subsubsection{Visibility or ``Showing Off''}
In Case B, we were intentional in designing opportunities for participants to regularly make their co-designed material visible to others in the workshop, such as through a walkthrough of their Problem Cards in the Lobby, a quick desk presentation to a co-participant of their action plan in the DIY Room, an open re-evaluation of their action plan in the Test Drive Room, and a concluding presentation in the Gallery. In Case A, as there were no co-participants in the workshop, the participants were excited to take the co-design material with them to ``show off'' or often share with their industry team members---a tangible reflection and dissemination of what they had build as a part of the workshop. This experience in Case A, reflected to a lesser degree than in Case B, also made us consider how we could design for participants to have a  ``Tangible Souvenir,'' discussed in detail in the next facet, ``Conceptual Achievement.''
\textbf{Visibility and ``Showing Off'' includes means for participants to present or otherwise share their work, thereby creating opportunities for reflection, engagement, and collaboration using the co-designed material as a platform for engagement within and/or across a group of participants.} As designers, we considered: \textit{How can participants share their work during multiple periods across the workshop? What different levels of visibility should we support, with different goals and types of desired feedback? How can we use visual metaphors to increase ludic engagement while sharing? What kinds of materials would participants be interested in sharing or ``showing off''?} 

\subsection{Conceptual Achievement}
Conceptual Achievement is a facet of co-design that supports participants' tangible and experiential sense of knowledge production, sharing, and evaluation, including: 1) framing intentions of the co-design; 2) alignment with participant expectations and mental models; and 3) balancing of various potential future-oriented layers of experiential residues. 

\begin{table*}[ht]
\centering
\caption{Facet: Conceptual Achievement}
\label{Codebook: Conceptual}
\Description[Table with two columns]{Table with two columns to present a list of sub-facets under the facet ``Conceptual Achievement.'' First column, ``Sub-facet'' lists different items under the main facets. Second column, ``Description'' defines each sub-facet. The rows of the table consists of five rows for each sub-facet described in detail in the paper and its description.}
\begin{tabularx}{\textwidth}{p{.25\textwidth}p{.7\textwidth}}
\toprule
\textbf{Sub-Facet} & \textbf{Description}\\ \midrule
\textbf{Performative Intentions} & \ldots refers to the intended outcomes and goals for the co-design workshop, marking both the starting point and potential teleology of co-design processes.\\

\textbf{Lived Experience as Expertise} & \ldots considers what knowledge participants bring with them and how they might use that knowledge to add to existing co-design materials, thereby aligning with and accomplishing the performative intentions.\\

\textbf{Mental Model Alignment} & \ldots considers how participants make sense of the co-design environment and performative intentions, and what kinds of interactions with the co-design space they believe are salient. \\

\textbf{Tangible Souvenirs} & \ldots include physical or digital artifacts that participants can take with them that were created through the workshop, thereby providing a sense of achievement and opportunity for sharing and dissemination. \\

\textbf{Layers of Experiential Residue} & \ldots considers potential impacts of the workshop on participants, including potential future activation of workshop goals, materials, and conversations. \\ 

\bottomrule
\end{tabularx}
\end{table*}

\subsubsection{Performative Intentions}
In both Cases A and B, we started our design process by drafting workshop goals that guided our entire design process of Material Engagement, Ludic Engagement, and Rhythms of Engagement. We stated the performative intentions for both cases in Sections \ref{CaseA} and \ref{CaseB}. 
In Case A, we intended for the participants to constantly reflect on and evaluate their current or past industry experience through the provided probes and reflective toolkits. In Case B, we intended for the participants to generate an action plan through the provided supports and building blocks to co-design their own action plan. In Case A, the performative intention was reflective and evaluative, whereas in Case B, it was generative; these differing intentions were reflected in chosen elements of the co-design space and material. 
\textbf{Performative Intentions refers to the intended outcomes and goals for the co-design workshop, marking both the starting point and potential teleology of co-design processes.} As designers, we asked: \textit{What do we want the participants to achieve through this co-design workshop? What constructively-focused learning outcomes do we hope for participants to attain? What do we need to include as part of the co-design experience for the participants to achieve these intentions?}

\subsubsection{Lived Experience as Expertise}
In both Cases A and B, we assumed the participant's knowledge as a primary mechanism through which they could meaningfully interact with and build upon the provided co-design materials. Aligning with Participatory Design approaches, we wanted our participants to bring their industry experience in their current professional role and constructively evaluate or generate new supports for their practice. In Case A, we sought for participants to use their existing knowledge of their practice to reflect on their own ecological complexity (Activity A), ethical dilemmas they had faced (Activity B), and the relevance of existing ethical supports in their work environment (Activity C). In Case B, beyond similar identification of ethical dilemmas and elements of ecological complexity, we also included an explicit space for the participants to add a tool that they already use in their current practice to the DIY Room options beyond those they had selected in the Shop. These constructive additions enabled the participants to have a sense of ownership, giving them ``permission'' to build on their own material(s) and knowledge. 
\textbf{Lived Experience as Expertise considers what knowledge participants bring with them and how they might use that knowledge to add to existing co-design materials, thereby aligning with and accomplishing the performative intentions.} As designers, we considered: \textit{What do we want participants to bring with them to the workshop? What expertise will enable participants to engage in sense-making with the provided materials? How can we support participants' ability to reshape or reframe elements of the workshop based on their lived experience?}

\subsubsection{Mental Model Alignment}
In Case B, we designed the materials to be accessible and easy for participants to engage in sense-making. We had to try out a range of visual metaphors that would help the participants relate to the performative intentions of each room in the metaphorical co-design ``house.'' For example, a Shop is a place where you collect items to eventually use them to build an action plan in the next stage (in DIY Room), and we relied on metaphors like ``shopping'' and ``do it yourself'' to indicate participants' ownership over the materials they selected and artifacts they created. In Case A, we sought to align participants' mental models through the material affordances of being able to map, filter, and annotate---however, this alignment was challenging for participants based on their ability to metacognitively engage in the activities and based on which activity was placed in which part of the sequence. For example, in Activity A, it was difficult for some of the participants to understand how visual mapping could represent their ecological complexity, requiring a lot of iteration in the toolkit design as well the scaffolds provided to the participants. Similarly, the mental model of evaluation for Activity C and the mental model of generation and speculation for Activity B required participants to both understand the instructions for a given activity, and then to know how to translate their experiences in relation to that activity structure. At an overall workshop level, we had to introduce the participants differently to the the Performative Intentions, one being reflective (Case A) and other being generative (Case B). In Case A, we provided worksheet material as probes for reflection from different perspectives which required participants to make sense of multiple independent mental models, whereas in Case B, the progression in the ``house'' provided them a sense of generating an action plan as part of one holistic mental model. 
\textbf{Mental Model Alignment considers how participants make sense of the co-design environment and performative intentions, and what kinds of interactions with the co-design space they believe are salient.} As designers, we had to constantly ask ourselves about the material designed: \textit{Will the participant understand the material provided and know how they are expected to interact? Do our performative intentions and activities align with mental models of a range of expected participant demographics? Do all activities relate to a single mental model or require participants' engagement with multiple discrete mental models?}

\subsubsection{Tangible Souvenirs}
In Case A, the idea of Tangible Souvenirs emerged when participants enthusiastically asked us if they could share the ecological maps (Activity A) and filtered ethical dilemmas (Activity B) they had created with their team members. The facilitator emailed them the co-designed material, which easily formed a packaged ``souvenir'' due to the ability of Miro to generate a multi-page PDF with one activity per page. In Case B, we intended the participants to apply their co-designed action plan in their everyday work and intentionally built the idea of a tangible outcome into the workshop sequence. We allowed Miro board access to the participants to refer back to their ``Problem Card'' (created in the Prep Room) and ``Action Plan'' (created in the DIY Room), and they had access to export and/or print their ethical action plans to share with other industry practitioners. 
\textbf{Tangible Souvenirs include physical or digital artifacts that participants can take with them that were created through the workshop, thereby providing a sense of achievement and opportunity for sharing and dissemination.} In our case, we intended the participants to use the artifacts they created in the workshop in their everyday practice for ethical decision making. As designers, we considered: \textit{How can we create artifacts that participants will want to share with others or their future self? What token can we provide to the participants as a remembrance of them sharing their expertise with the researchers? What outcomes could represent a pride of accomplishment for participants?} 

\subsubsection{Layers of Experiential Residue}
In both Cases A and B, we included a reflective session at the end of the workshop where we asked the participants to reflect on their learning (about themselves, their professional role, or their ecology) through the workshop. 
This reflection resulted in a range of answers which helped us evaluate and iterate upon our materials, while also becoming more intimately familiar with the intended and potential experiential factors that relate to the facets we detail in this paper. This feedback helped us to consider not only the designed experience, but also how this experience may contribute to \textit{attitudes} and \textit{opportunities for translational resonance} for participants during and after the workshop. 

\textbf{Layers of Experiential Residue considers potential impacts of the workshop on participants, including potential future activation of workshop goals, materials, and conversations.} As designers, we considered: \textit{How do we anticipate what attitudes or emotions we want participants to feel as they leave the workshop? How can we provide opportunities or ``hooks'' for potential future resonance of workshop activities with the participant's everyday work practices? What new actions or opportunities do we want participants to feel open to explore in their future work?} 
\begin{itemize}
    \item \textbf{Attitudes} consider the mood of participants both during and after participating in the workshop with \textit{emotions} such as happy, surprising, excited, tired, and exhausted. For example, most of the Case B participants were very intrigued and excited with the \textit{Visual Metaphors} provided through the Co-design Material and mentioned they had a different perception of a co-design space before our workshop. These attitudes also frame other emotions that might be related to the content of the workshop. In Cases A and B, we identified participants that had a sense of achievement and those that had a sense of failure. The \textit{sense of achievement} was our desired end goal, where participants learned through every step of the co-design workshop and achieving the intended outcomes, but often a felt imbalance of achievement from one step to another directly impacted the Emotional Residue. For example, in both Cases A and B, participants were reflecting on how thrilled and intrigued they were to have learnt about their own ecologies where they work everyday. This also impacted a few participants in making them realize the need of external supports leading them into dilemmas of how they could really achieve their action plans; although realization of such a fact was the first step. In contrast, participants that had a \textit{sense of failure} felt that they have not achieved what was asked of them or intended for them to experience. For example, in Case B, participants were expected to build an action plan by the end of the session---a tangible goal for the participants to achieve. Some participants felt a sense of ``failure'' due to their desire to build the ``perfect'' action plan for ethical decision-making for ``their'' own everyday practice. In response to these concerns, we edited our Session Script to provide participants assurance that there is ``\textit{no right or wrong way to make your action plan}'' as they are designing it for their own use in the future. It was important to constantly assure the participant that the co-design activities were meant to leverage their own ``lived'' experience and knowledge by treating them as the experts in the process, actively discouraging them from treat anything as if it was a ``failure.'' 
    \item \textbf{Opportunities for Translational Resonance} considers the conceptual after-effects that can be carried forward beyond the workshop into the everyday work or lives of participants, evoked through reflections, self-evaluation, and awareness built through the co-design experience. Opportunities could take the form of new awareness of action possibilities, new goals they wished to achieve, or new appreciation for complexity that was hidden to them before the workshop. These opportunities could then relate to potential resonance, with the participant assessing what part(s) of the experience remained relevant or salient after the participants leave the co-design space, including how it could impact their ``real life'', or in our case, their ``everyday practice.'' In both Cases A and B, participants identified their need of support for ethical action and carried forward their co-designed material as Tangible Souvenirs to share with their team members or as a self-awareness tool. Participants in both workshops also requested access to materials and in some cases found the potential for resonance in supporting future conversations with co-workers or other members of their profession. 
\end{itemize}

\section{``HATS'' WHILE DESIGNING FOR CO-DESIGN} \label{sec:hats}
Unlike the previous facets that focus on elements of sensitivity towards the specification or realization of co-design experiences, we derived various ``Hats'' that focused consideration on the different roles that people take on when constructing the co-design material and space (researchers and designers) and during the facilitation of the workshop (facilitators). 

 
\begin{table*}[ht]
\centering
\caption{``Hats'' that people take on when involved in co-design.}
\label{Codebook:Humanasfacilitator}
\Description[Table with two columns]{Table with two columns to present a list of roles taken during co-design. First column, ``Roles'' lists three roles such as ``Researchers,'' ``Designers,'' and ``Facilitators.'' Second column, ``Description'' defines each role's involvement in the design of co-design process. The rows of the table consists of three rows for each role described in detail in the paper and its description.}
\begin{tabularx}{\textwidth}{p{.25\textwidth}p{.7\textwidth}}
\toprule
\textbf{Roles} & \textbf{Description}\\ 
\midrule
\textbf{Researchers} & \ldots set the workshop goals and provide a design frame and constraints for the designers to design the co-design material, focusing on identification of relevant supports for conceptual achievement through the process. \\

\textbf{Designers} & \ldots design the co-design material and are constant evaluators of rhythm of engagement, material engagement, and ludic engagement, considered through the lens of conceptual achievement.\\

\textbf{Facilitators} & \ldots engage, motivate, and direct participants in the co-design workshop, supporting designed elements of the experience (e.g., rhythm, material, and ludic engagement) while also encouraging participants to identify their own unique path through the activities that have been prepared by researchers and designers. \\ 

\bottomrule
\end{tabularx}
\end{table*}

\subsubsection{Designers and Researchers}
The two main roles prevalent in the design of the co-design material and space are researchers and designers. While intertwined in practice, \textit{Researchers} set the workshop goals (with a focus on \textit{Performative Intentions}) and provide a frame and guiding set of constraints for the designers to design the co-design materials, with a primary focus on how \textit{Conceptual Achievement} relates to and can be supported by research goals. \textit{Designers} design the co-design material and are constant evaluators of the \textit{Rhythm of Engagement, Material, and Ludic Engagement}, considered through the framing of \textit{Conceptual Achievement}. Researchers and designers may represent the same individual, while taking on different and distinct roles with a difference in directionality and facet(s) considered. In Case A, the first author had to wear a range of hats being the researcher, designer, practitioner, and facilitator of the co-design workshops. In Case B, we had a group of researchers and designers (including the first and second author) who often more clearly divided the roles taken on. For example, as part of a group of designers, four undergraduate researchers mainly took on the role of designing the visual metaphors and tangible co-design material, maintaining conversation with the whole group and led by the first and second author, who primarily took on the role of researcher. Through conversation, the design of co-design materials was often realigned or adjusted based on aspects of conceptual achievement desired by the researchers. Across both Cases A and B, we also had to consider the ``practitioner hat'' to constantly relate our design decisions to the potential experiences of technology and design industry practitioners based on our own internship or full-time industry experiences. This constant and reflexive engagement with different roles allowed us to design our material and space to be more relevant and accessible to our participants---while also shifting between researcher and designer roles based on which constraints were in flux or under consideration. 

\subsubsection{Facilitators}
The facilitator role emerges in concrete form later in the process, but must nevertheless be considered throughout the design process. \textit{Facilitators} take on the role of engaging, motivating, and directing the participants in the co-design workshops. In a recent article, Dahl and Sharma \cite{Dahl2022-fg} listed six facets of the facilitator role: (1) trust builder, (2) enabler, (3) inquirer, (4) direction setter, (5) value provider, and (6) users' advocate. 
In addition to these elements of the facilitator role, the facilitators have to take care of the logistics ``in the moment'' including time-keeping, recording, music, grouping, session scripts, providing participant guidance, and improvising when participants identify new and unintended pathways through the workshop activities. In Case B, given that we were designing for multiple participants through a digital environment that required multiple facilitators, we further categorized the roles of facilitators as:

\begin{itemize}
    \item \textbf{Main facilitator:} A facilitator who ensures the overall functioning of the workshop, including holistic engagement with key rhythms and forms of material and ludic engagement. In our case, the main facilitator was in the main Zoom Room to align all the other facilitators with the overall timing for the workshop, managed the Slack behind-the-scenes discussions to align the co-facilitators, and developed plans for improvisation based on the experiences of co-facilitators.
    \item \textbf{Co-Facilitator:} A facilitator who is responsible for a subset of responsibilities, activities, or number of participants, working in parallel with other co-facilitators to ensure the success of the workshop. In our case, a co-facilitator managed each sub-group in a breakout room, including responsibility for all subgroup logistics such as script, recording, facilitation, and contributions to the Slack backchannel communication with the main facilitator.
\end{itemize}

The organization of these facilitators enabled specific categorization and assignment of tasks allowing for a smooth and orchestrated Rhythm of Engagement. In addition to these aspects of facilitation and the dimensions proposed by Dahl and Sharma~\cite{Dahl2022-fg}, we also encountered additional faciliation factors that encourage consideration of emotional, identity, and attitudinal characteristics. Facilitators are indirectly the building blocks for excitement and curiosity in a workshop, building upon the presence of active or calm participants, or some combination of both. A facilitator's \textit{attitudinal characteristics} may contribute to directing the energy of enthusiastic participants, elevating the energy of calm and initially awkward participants, and rejuvenating the energy of participants as they engage for longer periods of time in the workshop. These attributes of facilitation also relate to the \textit{identity characteristics} of the facilitator, who often has to think on their feet regarding how to \textit{improve} the engagement among participants as required through management of rhythms and ludic engagement, including opportunities for potential improvisation to improve workshop and participant outcomes---all while building on prior experiences and identity commitments that enable them to empathize with and understand different participant profiles. Finally, a facilitator's \textit{emotional characteristics} may contribute to their success or help to explain their fatigue. Facilitators often have to exert higher amounts of energy compared to anyone else in a workshop, seamlessly handling the Rhythm of Engagement, Material Engagement, Entertainment, Conceptual Engagement of/by the participants, Session Scripts, Zoom room switching and transitions, Miro Board interactions, and recording at every transition, which demands them to be verbally and mentally active at all times. This extensive engagement can lead to fatigue in ways that can impacts their attitudinal characteristics and ability to engage in a personable manner over an extended period of time---particularly in digital formats such as those we designed in Cases A and B. 

\section{Discussion and Implications}
In the previous sections, we have laid the groundwork for the formation of new intermediate-level knowledge that can support the designers of co-design experiences, consisting of four facets: Rhythm of Engagement, Material Engagement, Ludic Engagement, and Conceptual Achievement, all framed through three ``hats'' or roles of designer, researcher, and facilitator. These facets were formed through analysis of various design decisions that we made to create two different co-design workshop structures, linking existing best practices and principles to concrete design processes, resulting in a set of intermediate-level knowledge that includes a foundation for patterns, heuristics, and strong concepts to support future co-design practices. In this section, we further discuss these forms of knowledge which can be activated or supported using these facets and roles individually, or in combination. First, we illustrate the interactional qualities of these facets that might encourage different attitudes towards other forms of knowledge that are leveraged when designing for co-design. Second, we describe how this range of facets and sub-facets can be framed as differing forms of intermediate-level knowledge, laying the groundwork for further research and targeted knowledge production in the co-design literature. 

\subsection{Interactional Qualities of Facets} \label{sec:interactions}

Each facet we have described above forms a useful set of knowledge that co-design researchers and designers could engage with during the design of a co-design experience. We also propose that the utility of these facets may exist beyond the facets as individual entities, and may also include non-exclusive, interlinked, or inter-dependent qualities of these facets. For instance, the use of various groupings of facets as different sets of intermediate-level knowledge may illuminate or otherwise support different aspects of designing for co-design. While our goal is not to identify all such groupings, we link our findings to principles from existing co-design literature to illustrate how these facets may productively extend and support existing principal knowledge. 

\begin{figure}[ht]
    \centering
    \subfigure[]{\includegraphics[width=0.45\textwidth]{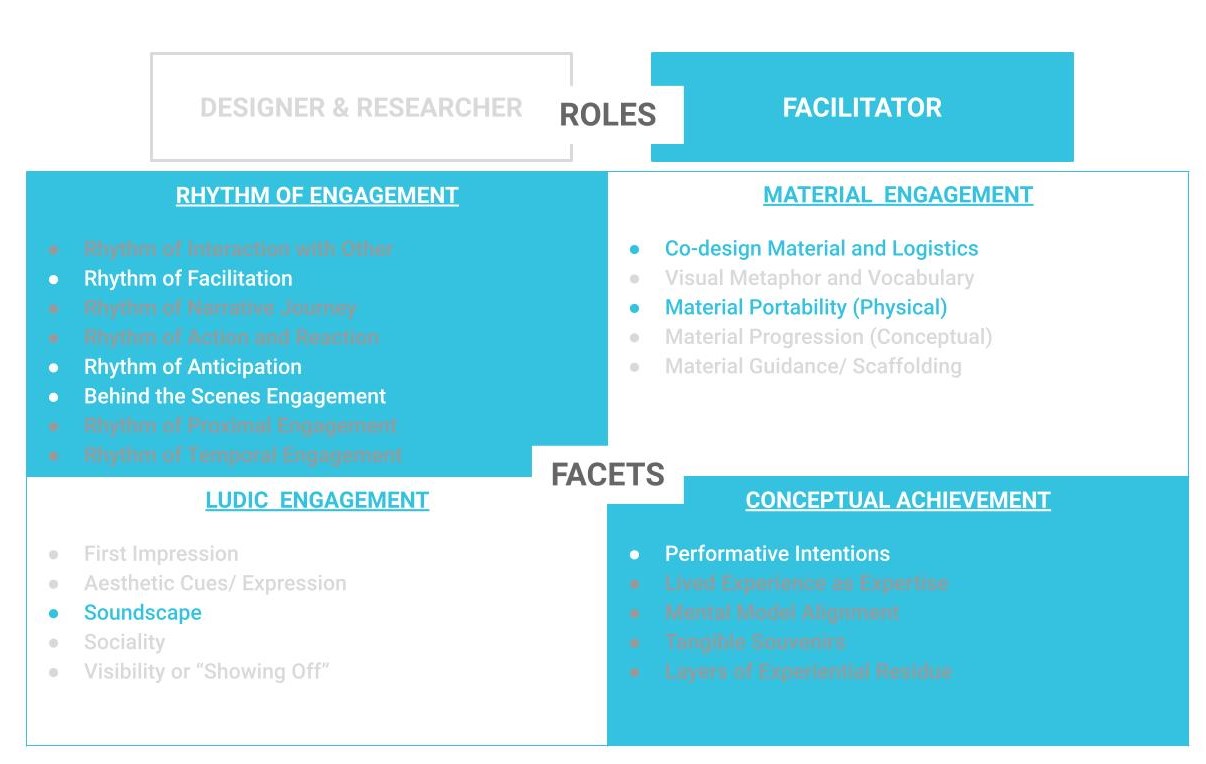}} 
    \subfigure[]{\includegraphics[width=0.45\textwidth]{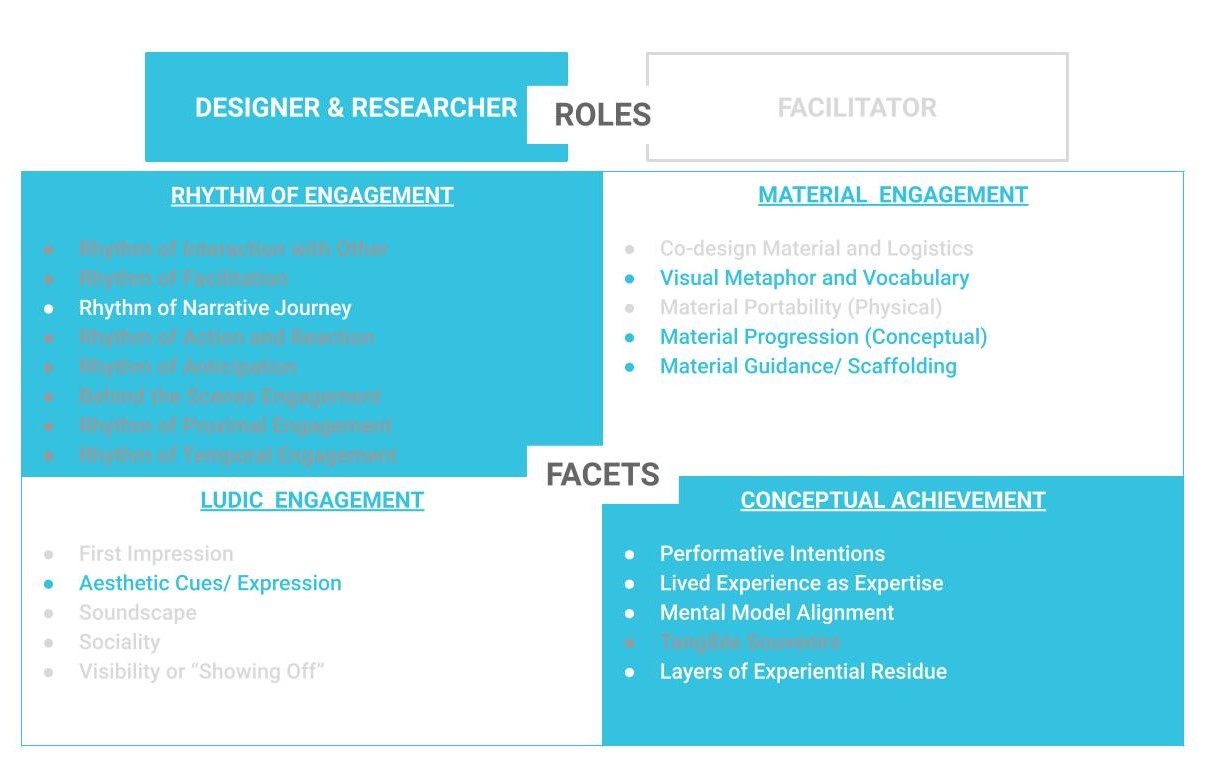}} 
    \subfigure[]{\includegraphics[width=0.8\textwidth]{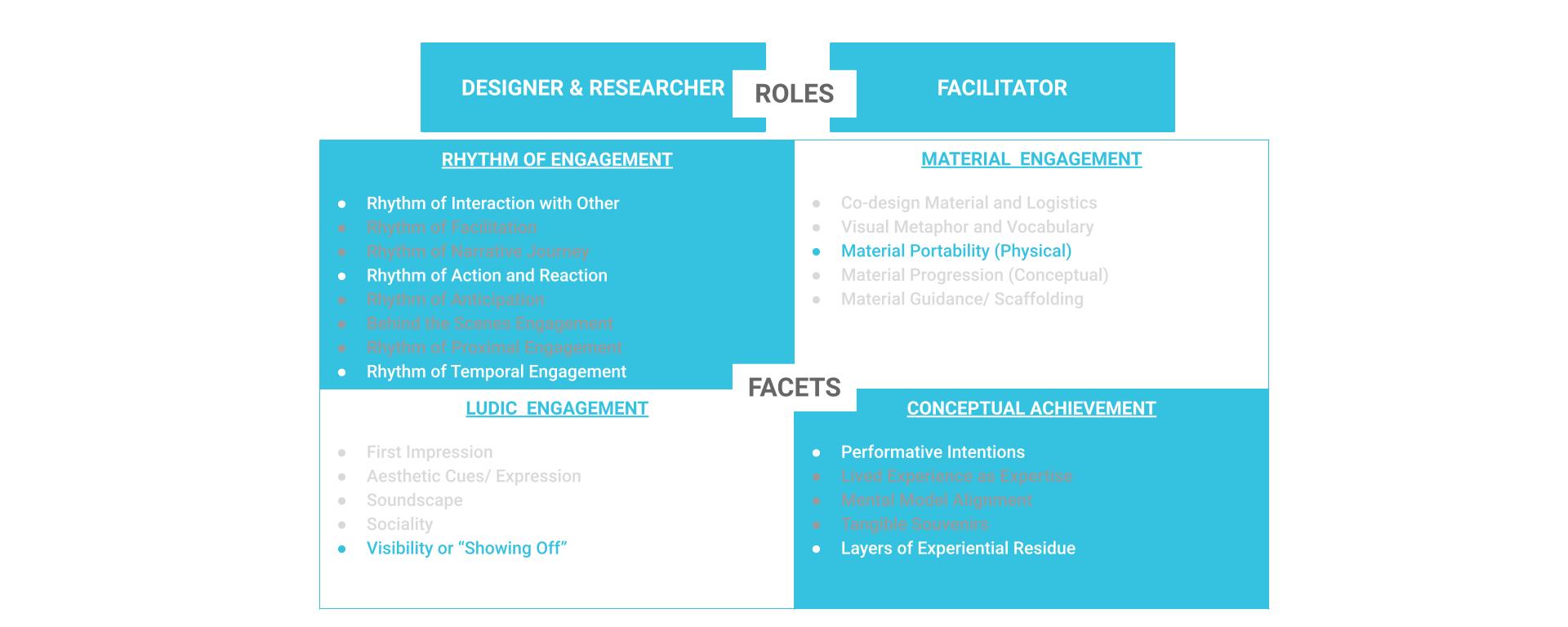}}
    \caption{Interactional qualities of Facets and Roles to describe different potential sets of design knowledge.}
    \label{fig:IQ1-3}
    \Description[Figure with three sub-figures]{Figure consists of three sub-figures named (a), (b), and (c). Each sub-figure reuses the schema in Figure~\ref{fig:facets}. In each sub-figure, different listed sub-facets text is struck out through lighter text color to de-highlight them compared to others to work as a visual reference to the interactional qualities discussed in the text following the Figure. For example, in sub-figure (a), role ``Facilitator'' rectangle is highlighted in blue rectangle and ``Designer \& Researcher'' rectangle is greyed out. Same follows for the text of the sub-facets listed in the 2X2 matrix.}
\end{figure}

\begin{itemize}
    \item \textit{Building Facilitation Capacity:} The facilitator role, as described by Dahl and Sharma~\cite{Dahl2022-fg}, involves being a trust builder, enabler, inquirer, direction setter, value provider, and users' advocate. To further support these roles, we can identify an interaction among facets that supports a characterization of facilitators' knowledge that enables participation through intentional design of the co-design experience and material. As shown in Figure~\ref{fig:IQ1-3}(a), designers can: 1) provide \textit{direction} or be an \textit{enabler} by considering the Rhythm of Facilitation (sub-facet of Rhythm of Engagement) in which they anticipate how they might guide and intervene during participants' engagement with the material, managing Material Portability (sub-facet of Material Engagement) or Soundscape (sub-facet of Ludic Engagement); 2) ensure that co-design participants derive \textit{value} by considering how they can achieve meaningful Performative Intentions (sub-facet of Conceptual Achievement), supported by Behind-the-Scenes Engagement (sub-facet of Rhythm of Engagement) to align with co-facilitators at times of Rhythm of Anticipation (sub-facet of Rhythm of Engagement); and 3) build mechanisms for \textit{user advocacy} and \textit{trust building} into a detailed \textit{Session Script} (Co-design Material and Logistics, a sub-facet of Material Engagement), which considers what might feel natural or jarring to a participant. 
    
    \item \textit{Scaffolding for Perceived Creativity:} One goal of co-design experiences as outlined by Sanders and Stappers~\cite{Sanders2008-eq} is to create experiences that allow participants with many different levels of perceived creativity (from ``doing'' to ``making'' to ``creating'') to engage meaningfully. To support consideration of these levels, designers might consider a range of facets to create effective scaffolds through the co-design space and material for participants to achieve the workshop goals. As shown in Figure~\ref{fig:IQ1-3}(b), designers can: 1) consider the creation of Scaffolds or Guidance through Material (sub-facet of Material Engagement) by identifying relatable and appropriate Visual Metaphors and Vocabulary (another sub-facet of Material Engagement) that align with the Mental Model (sub-facet of Conceptual Achievement) of the participants; 2) identify a range of Aesthetic Cues (sub-facet of Ludic Engagement) and Material Progression (sub-facet of Material Engagement) that allows participants at multiple levels of creativity to meaningfully and playfully express themselves; and 3) create meaningful goals for participants that are expressed through understandable Performative Intentions that result in a positive Experiential Residue (sub-facets of Conceptual Achievement) to allow the participant feel that their co-design experience was personally or professionally meaningful. 
    
    \item \textit{Sharing Co-designed Knowledge through Collective Creativity:} Another goal of co-design experiences outlined by Sanders and Stappers~\cite{Sanders2008-eq} is the creation of an environment that can be characterized by collaboration, generativity, and a collective sense of creativity. To support the creation of a collectively creative space, designers might consider a range of facets for collaboration through action, reflection, and evaluation of co-designed knowledge, while also activating the facilitator's knowledge in encouraging collective sharing of this knowledge (activating two roles as once). As shown in Figure~\ref{fig:IQ1-3}(c), designers can: 1) create opportunities for participants to share their produced material with other participants in the workshop through a Rhythm of Interaction with Others in smaller groups and Rhythm of Action and Reaction with the co-designed material (sub-facets of Rhythm of Engagement) in a way that is not overwhelming for the participants while also providing facilitator support for Material Portability (sub-facet of Material Engagement); 2) balance the Rhythm of Temporal Engagement effectively so it does not disrupt the progression of the Performative Intentions (sub-facet of Conceptual Achievement), regulating participation between the end states of not sharing or ``oversharing'' (with Rhythm of Temporal Engagement); 3) provide opportunities for participants to ``Show Off'' (sub-facet of Ludic Engagement) their knowledge with others and provide support for appropriate collective reactions from others in the workshop---feedback, appreciation, and/or reflection (i.e., Rhythm of Action and Reaction); and finally, 4) identify relevant Layers of Experiential Residue (sub-facet of Conceptual Achievement) that the participant should exit the experience, especially due to the presence and interaction with fellow co-designers, with using the previous facet interactions as inspiration. 
    
\end{itemize}

\subsection{Towards an Intermediate-Level Knowledge of Co-Design}
Building on our findings in this paper, we identify opportunities for future work that could build other forms of knowledge that extend our work. We focus our attention on the potential of fostering intermediate-level knowledge generation practices to strengthen the foundations of co-design practices.

As showcased in Figure~\ref{fig:ILK}, we have constructed, reflected upon, and described the interplay of particular forms of intermediate-level knowledge through the four facets and three roles---linking to types such as heuristics, patterns, and strong concepts. As proposed by L\"owgren and H\"o\"ok~\cite{Lowgren2013-db,Hook2012-dd}, there are various other kinds and forms of knowledge at different levels between ``theory'' and specific cases, which in the context of co-design might include knowledge forms such as methods and tools, patterns, experiential qualities, annotated portfolios, criticism, and concepts. Leveraging the knowledge contribution of this paper, we call for explicit knowledge generation in relation to a range of co-design practices that supports these intermediate-knowledge types, thereby better supporting co-design and co-creation practices in HCI and design research. This knowledge generation process could address multiple different goals, including engaging a broader range of research methods; identifying and utilizing different sets standards to judge knowledge contributions based on the standards of quality relevant for a particular type of intermediate-level knowledge (e.g., what makes something a ``rigorous'' principle may be different from what makes something a ``useful'' pattern); and articulating connections between various forms of knowledge and the design of co-design experiences. 

First, we encourage further investigation of co-design knowledge using different research methods, alongside co/auto-ethnography, thereby developing pragmatic and rigorous accounts of designing for co-design. For instance, established co-designers might consider creating curated sets of cases around co-design that illustrate a range of topic domains and contexts in which the methodology is successfully used. Similarly, in-depth descriptions of the \textit{experiential qualities} relating to co-design work may allow scholars to elevate the results of participatory approaches of the participants engaging in such practices within HCI and design communities. A consideration of the range of appropriate practices will be important to consider, likely ranging from design-focused methodologies such as Research through Design to evaluation-focused methods such as those commonly used for method validation to formal empirical work. Across these inquiry approaches, core questions should be considered that relate to design knowledge, the status of participation, the leveraging of experiential qualities, and the political and ontological status of co-design as a practice. 

Second, our contribution to the intermediate-level knowledge space begs the question: how can intermediate-level knowledge be formed, how can its quality be judged (and by what standards), and how do we know when we have enough knowledge in a particular area? While we cannot answer this question in full, co-design scholars and practitioners may consider how theoretical or principial knowledge of co-design might be framed and judged differently from reports on specific instances of co-design. For instance, building on the development of the \textit{design case} as a specific form of knowledge building in design~\cite{Boling2012-od,Gray2020-yu} which draws on traditions of utility rather than generalizability, may be instructive in the generation and curation of intermediate-level knowledge to support co-design practices. In our work, we have proposed an initial set of facets and roles which may be expanded into sets of patterns (e.g., what are common patterns of tangible souvenirs that can be used to document conceptual achievement), strong concepts (e.g., what does it mean to ludic-ly engage in a co-design session), or annotated portfolios (e.g., how can we juxtapose different types of rhythms that create a different feeling of progression?). The vocabulary we have proposed in this paper may also support the framing of existing co-design examples and/or generation of more detailed accounts of the design of co-design experiences, encouraging sensitization on the part of designers or design teams towards specific---and perhaps unacknowledged---parts of their design work that could have value to other designers. The vocabulary we provided may also be used as an initial structure for co-design practitioners to re-evaluate about their de-prioritization of certain ``facets,'' either in their past or current co-design practice, impacts the outcomes of their workshops. This awareness, in turn, may contribute to intermediate-level knowledge based on evaluative qualities as realized in use.  

Third, the expansion of \textit{methods and tools} may support a realignment of co-design scholarship, supporting an explicit conversation regarding the ``how'' aspect of decision making while designing for co-design rather than primarily the ``what'' of the co-design. The intermediate-level knowledge generated through this reframing may enable HCI and design researchers to more fully describe \textit{when} to use co-design as a methodology, \textit{how} to frame identify various tools and methods used to operationalize co-design methodology as an extension to designer's capabilities, and \textit{what} to provide in the spectrum of toolkits, probes, and/or activities based on the co-design space, material, or intention. Further, future scholarship may further demonstrate how co-design materials can be used as primitives to support the explicit design of methods for repetitive and generative use across a range of co-design spaces and contexts.

\section{Conclusion}
In this paper, we present outcomes of our co/auto-ethnographic efforts, describing new intermediate-level knowledge that co-design researchers and designers might use to design for co-design workshops. We describe four multi-dimensional facets (Rhythm of Engagement, Material Engagement, Ludic Engagement, and Conceptual Achievement) and three ``hats'' or roles involved (researchers, designers, and facilitators) in the design of co-design experiences. We use these findings to describe a landscape of \textit{intermediate-level knowledge} used by co-design researchers and designers in their \textit{designerly practice} as they consider the co-design environment, material, and intentions. Through the identified facets and roles (in Section~\ref{sec:interactions}), we identify the interactive qualities and non-exclusive nature of these facets, underscoring designers' use of a range of knowledge forms during the design of co-design experiences and calling for additional attention to a range of knowledge generation practices to support these designers. 


\begin{acks}
This work is funded in part by the National Science Foundation under Grant No. 1909714. We thank and  gratefully acknowledge the efforts of graduate researchers Ziquing Li, Ike Obi; and undergraduate researchers Anne Pivonka, Ambika Menon, Matthew Will, Janna Johns, and Thomas Carlock for their contributions to Case B design process. 
\end{acks}

\bibliographystyle{ACM-Reference-Format}
\bibliography{meta}


\begin{thebibliography}{69}


\ifx \showCODEN    \undefined \def \showCODEN     #1{\unskip}     \fi
\ifx \showDOI      \undefined \def \showDOI       #1{#1}\fi
\ifx \showISBNx    \undefined \def \showISBNx     #1{\unskip}     \fi
\ifx \showISBNxiii \undefined \def \showISBNxiii  #1{\unskip}     \fi
\ifx \showISSN     \undefined \def \showISSN      #1{\unskip}     \fi
\ifx \showLCCN     \undefined \def \showLCCN      #1{\unskip}     \fi
\ifx \shownote     \undefined \def \shownote      #1{#1}          \fi
\ifx \showarticletitle \undefined \def \showarticletitle #1{#1}   \fi
\ifx \showURL      \undefined \def \showURL       {\relax}        \fi
\providecommand\bibfield[2]{#2}
\providecommand\bibinfo[2]{#2}
\providecommand\natexlab[1]{#1}
\providecommand\showeprint[2][]{arXiv:#2}

\bibitem[Akama et~al\mbox{.}(2013)]%
        {Akama2013-pb}
\bibfield{author}{\bibinfo{person}{Yoko Akama}, \bibinfo{person}{Alison
  Prendiville}, {and} \bibinfo{person}{{Others}}.}
  \bibinfo{year}{2013}\natexlab{}.
\newblock \showarticletitle{Embodying, enacting and entangling design: A
  phenomenological view to co-designing services}.
\newblock \bibinfo{journal}{\emph{Swedish Design Research Journal}}
  \bibinfo{volume}{1}, \bibinfo{number}{1} (\bibinfo{year}{2013}),
  \bibinfo{pages}{29--41}.
\newblock


\bibitem[Blomkamp(2018)]%
        {Blomkamp2018-iv}
\bibfield{author}{\bibinfo{person}{Emma Blomkamp}.}
  \bibinfo{year}{2018}\natexlab{}.
\newblock \bibinfo{title}{Sharing the principles of co-design}.
\newblock
  \bibinfo{howpublished}{\url{https://medium.com/@emmablomkamp/sharing-the-principles-of-co-design-4a976bb55c48}}.
\newblock
\newblock
\shownote{Accessed: 2020-2-21}.


\bibitem[Boling et~al\mbox{.}(2020)]%
        {Boling2020-ci}
\bibfield{author}{\bibinfo{person}{Elizabeth Boling}, \bibinfo{person}{Colin~M
  Gray}, {and} \bibinfo{person}{Kennon~M Smith}.}
  \bibinfo{year}{2020}\natexlab{}.
\newblock \showarticletitle{Educating for design character in higher education:
  Challenges in studio pedagogy}. In \bibinfo{booktitle}{\emph{Proceedings of
  the Design Research Society}} (Brisbane, Australia).
  \bibinfo{publisher}{Design Research Society}.
\newblock
\urldef\tempurl%
\url{https://doi.org/10.21606/drs.2020.120}
\showDOI{\tempurl}


\bibitem[Boling and Smith(2012)]%
        {Boling2012-od}
\bibfield{author}{\bibinfo{person}{Elizabeth Boling} {and}
  \bibinfo{person}{Kennon~M Smith}.} \bibinfo{year}{2012}\natexlab{}.
\newblock \showarticletitle{{The design case}}.
\newblock \bibinfo{journal}{\emph{Interactions}} \bibinfo{volume}{19},
  \bibinfo{number}{5} (\bibinfo{date}{Sept.} \bibinfo{year}{2012}),
  \bibinfo{pages}{48--53}.
\newblock
\showISSN{1072-5520, 1558-3449}
\urldef\tempurl%
\url{https://doi.org/10.1145/2334184.2334196}
\showDOI{\tempurl}


\bibitem[Christensen et~al\mbox{.}(2017)]%
        {Christensen2017-tm}
\bibfield{author}{\bibinfo{person}{Bo~T Christensen}, \bibinfo{person}{Linden~J
  Ball}, {and} \bibinfo{person}{Kim Halskov}.} \bibinfo{year}{2017}\natexlab{}.
\newblock \bibinfo{booktitle}{\emph{Analysing Design Thinking: Studies of
  {Cross-Cultural} {Co-Creation}}}.
\newblock \bibinfo{publisher}{CRC Press}.
\newblock
\showISBNx{9781351802833}
\urldef\tempurl%
\url{https://play.google.com/store/books/details?id=bjkPEAAAQBAJ}
\showURL{%
\tempurl}


\bibitem[Coia and Taylor(2009)]%
        {Coia2009-bf}
\bibfield{author}{\bibinfo{person}{Lesley Coia} {and} \bibinfo{person}{Monica
  Taylor}.} \bibinfo{year}{2009}\natexlab{}.
\newblock \showarticletitle{Co/autoethnography: Exploring Our Teaching Selves
  Collaboratively}.
\newblock In \bibinfo{booktitle}{\emph{Research Methods for the Self-study of
  Practice}}, \bibfield{editor}{\bibinfo{person}{Linda Fitzgerald},
  \bibinfo{person}{Melissa Heston}, {and} \bibinfo{person}{Deborah Tidwell}}
  (Eds.). \bibinfo{publisher}{Springer Netherlands},
  \bibinfo{address}{Dordrecht}, \bibinfo{pages}{3--16}.
\newblock
\showISBNx{9781402095146}
\urldef\tempurl%
\url{https://doi.org/10.1007/978-1-4020-9514-6\_1}
\showDOI{\tempurl}


\bibitem[Costanza-Chock(2020)]%
        {Costanza-Chock2020-vf}
\bibfield{author}{\bibinfo{person}{Sasha Costanza-Chock}.}
  \bibinfo{year}{2020}\natexlab{}.
\newblock \bibinfo{booktitle}{\emph{Design Justice: {Community-Led} Practices
  to Build the Worlds We Need}}.
\newblock \bibinfo{publisher}{MIT Press}.
\newblock
\showISBNx{9780262356879}


\bibitem[Cottam and Leadbeater(2004)]%
        {Cottam2004-di}
\bibfield{author}{\bibinfo{person}{Hilary Cottam} {and}
  \bibinfo{person}{Charles Leadbeater}.} \bibinfo{year}{2004}\natexlab{}.
\newblock \showarticletitle{{RED} paper 01: Health: Co-creating services}.
\newblock \bibinfo{journal}{\emph{London: Design Council}}
  (\bibinfo{year}{2004}).
\newblock


\bibitem[Cross(2001)]%
        {Cross2001-zg}
\bibfield{author}{\bibinfo{person}{Nigel Cross}.}
  \bibinfo{year}{2001}\natexlab{}.
\newblock \showarticletitle{{Designerly Ways of Knowing: Design Discipline
  Versus Design Science}}.
\newblock \bibinfo{journal}{\emph{Design Issues}} \bibinfo{volume}{17},
  \bibinfo{number}{3} (\bibinfo{date}{Jan.} \bibinfo{year}{2001}),
  \bibinfo{pages}{49--55}.
\newblock
\showISSN{0747-9360}
\urldef\tempurl%
\url{https://doi.org/10.1162/074793601750357196}
\showDOI{\tempurl}


\bibitem[Cross(2007)]%
        {Cross2007-ei}
\bibfield{author}{\bibinfo{person}{N Cross}.} \bibinfo{year}{2007}\natexlab{}.
\newblock \showarticletitle{From a design science to a design discipline:
  Understanding designerly ways of knowing and thinking}.
\newblock In \bibinfo{booktitle}{\emph{Design Research Now: Essays and Selected
  Projects}}, \bibfield{editor}{\bibinfo{person}{Ralf Michel}} (Ed.).
  \bibinfo{publisher}{Birkh{\"a}user}, \bibinfo{address}{Basel, Switzerland},
  \bibinfo{pages}{41--54}.
\newblock


\bibitem[Cruickshank et~al\mbox{.}(2013)]%
        {Cruickshank2013-af}
\bibfield{author}{\bibinfo{person}{Leon Cruickshank}, \bibinfo{person}{Gemma
  Coupe}, {and} \bibinfo{person}{Dee Hennesy}.}
  \bibinfo{year}{2013}\natexlab{}.
\newblock \showarticletitle{Beyond the castle:public space co-design, a case
  study and guidelines for designers}.
\newblock \bibinfo{journal}{\emph{Swedish Design Research Journal}}
  \bibinfo{volume}{2} (\bibinfo{year}{2013}), \bibinfo{pages}{10}.
\newblock
\showISSN{2000-7574}


\bibitem[Dahl and Sharma(2022)]%
        {Dahl2022-fg}
\bibfield{author}{\bibinfo{person}{Yngve Dahl} {and} \bibinfo{person}{Kshitij
  Sharma}.} \bibinfo{year}{2022}\natexlab{}.
\newblock \showarticletitle{Six Facets of Facilitation: Participatory Design
  Facilitators' Perspectives on Their Role and Its Realization}. In
  \bibinfo{booktitle}{\emph{{CHI} Conference on Human Factors in Computing
  Systems}} (New Orleans, LA, USA) \emph{(\bibinfo{series}{CHI '22},
  \bibinfo{number}{Article 484})}. \bibinfo{publisher}{Association for
  Computing Machinery}, \bibinfo{address}{New York, NY, USA},
  \bibinfo{pages}{1--14}.
\newblock
\showISBNx{9781450391573}
\urldef\tempurl%
\url{https://doi.org/10.1145/3491102.3502013}
\showDOI{\tempurl}


\bibitem[Dede et~al\mbox{.}(2012)]%
        {Dede2012-mh}
\bibfield{author}{\bibinfo{person}{Okan~Murat Dede}, \bibinfo{person}{{\c
  C}igdem~Belgin Dikmen}, {and} \bibinfo{person}{Asim~Mustafa Ayten}.}
  \bibinfo{year}{2012}\natexlab{}.
\newblock \showarticletitle{A new approach for participative urban design: An
  urban design study of Cumhuriyet urban square in Yozgat Turkey}.
\newblock \bibinfo{journal}{\emph{Journal of Geography and Regional Planning}}
  \bibinfo{volume}{5}, \bibinfo{number}{5} (\bibinfo{year}{2012}),
  \bibinfo{pages}{122}.
\newblock


\bibitem[Del~Gaudio et~al\mbox{.}(2018)]%
        {Del_Gaudio2018-zi}
\bibfield{author}{\bibinfo{person}{Chiara Del~Gaudio}, \bibinfo{person}{Carlo
  Franzato}, {and} \bibinfo{person}{Alfredo~Jefferson de Oliveira}.}
  \bibinfo{year}{2018}\natexlab{}.
\newblock \showarticletitle{Co-design for democratising and its risks for
  democracy}.
\newblock \bibinfo{journal}{\emph{CoDesign}} (\bibinfo{date}{Dec.}
  \bibinfo{year}{2018}), \bibinfo{pages}{1--18}.
\newblock
\showISSN{1571-0882}
\urldef\tempurl%
\url{https://doi.org/10.1080/15710882.2018.1557693}
\showDOI{\tempurl}


\bibitem[Dunne(1999)]%
        {Dunne1999-ub}
\bibfield{author}{\bibinfo{person}{Joseph Dunne}.}
  \bibinfo{year}{1999}\natexlab{}.
\newblock \showarticletitle{{Professional judgment and the predicaments of
  practice}}.
\newblock \bibinfo{journal}{\emph{European journal of marketing}}
  \bibinfo{volume}{33}, \bibinfo{number}{7/8} (\bibinfo{date}{Jan.}
  \bibinfo{year}{1999}), \bibinfo{pages}{707--720}.
\newblock
\showISSN{0309-0566}


\bibitem[Durall et~al\mbox{.}(2020)]%
        {Durall2020-zz}
\bibfield{author}{\bibinfo{person}{Eva Durall}, \bibinfo{person}{Merja
  Bauters}, \bibinfo{person}{Iida Hietala}, \bibinfo{person}{Teemu Leinonen},
  {and} \bibinfo{person}{Evangelos Kapros}.} \bibinfo{year}{2020}\natexlab{}.
\newblock \showarticletitle{Co-creation and co-design in technology-enhanced
  learning: Innovating science learning outside the classroom}.
\newblock \bibinfo{journal}{\emph{Interaction Design and Architecture (s)}}
  \bibinfo{volume}{42} (\bibinfo{year}{2020}), \bibinfo{pages}{202--226}.
\newblock


\bibitem[Evans and Terrey(2016)]%
        {Evans2016-lr}
\bibfield{author}{\bibinfo{person}{Mark Evans} {and} \bibinfo{person}{Nina
  Terrey}.} \bibinfo{year}{2016}\natexlab{}.
\newblock \showarticletitle{Co-design with citizens and stakeholders}.
\newblock \bibinfo{journal}{\emph{Evidence-based Policymaking in the Social
  Sciences: Methods that Matter}} (\bibinfo{year}{2016}),
  \bibinfo{pages}{243--262}.
\newblock


\bibitem[Fitton et~al\mbox{.}(2018)]%
        {Fitton2018-cw}
\bibfield{author}{\bibinfo{person}{Daniel Fitton}, \bibinfo{person}{Janet~C
  Read}, \bibinfo{person}{Gavin Sim}, {and} \bibinfo{person}{Brendan Cassidy}.}
  \bibinfo{year}{2018}\natexlab{}.
\newblock \showarticletitle{Co-designing Voice User Interfaces with Teenagers
  in the Context of Smart Homes}. In \bibinfo{booktitle}{\emph{Proceedings of
  the 17th {ACM} Conference on Interaction Design and Children}} (Trondheim,
  Norway) \emph{(\bibinfo{series}{IDC '18})}. \bibinfo{publisher}{ACM},
  \bibinfo{address}{New York, NY, USA}, \bibinfo{pages}{55--66}.
\newblock
\showISBNx{9781450351522}
\urldef\tempurl%
\url{https://doi.org/10.1145/3202185.3202744}
\showDOI{\tempurl}


\bibitem[Gaver et~al\mbox{.}(1999)]%
        {Gaver1999-fi}
\bibfield{author}{\bibinfo{person}{Bill Gaver}, \bibinfo{person}{Tony Dunne},
  {and} \bibinfo{person}{Elena Pacenti}.} \bibinfo{year}{1999}\natexlab{}.
\newblock \showarticletitle{Design: cultural probes}.
\newblock \bibinfo{journal}{\emph{Interactions}} \bibinfo{volume}{6},
  \bibinfo{number}{1} (\bibinfo{year}{1999}), \bibinfo{pages}{21--29}.
\newblock
\showISSN{1072-5520}


\bibitem[Gispen(2017)]%
        {ethicalcontract}
\bibfield{author}{\bibinfo{person}{Jet Gispen}.}
  \bibinfo{year}{2017}\natexlab{}.
\newblock \bibinfo{title}{Ethical Contract}.
\newblock
  \bibinfo{howpublished}{\url{https://www.ethicsfordesigners.com/ethical-contract}}.
\newblock
\newblock
\shownote{Accessed: 2020-5-12}.


\bibitem[Goodman et~al\mbox{.}(2011)]%
        {Goodman2011-ak}
\bibfield{author}{\bibinfo{person}{Elizabeth Goodman}, \bibinfo{person}{Erik
  Stolterman}, {and} \bibinfo{person}{Ron Wakkary}.}
  \bibinfo{year}{2011}\natexlab{}.
\newblock \showarticletitle{Understanding Interaction Design Practices}. In
  \bibinfo{booktitle}{\emph{Proceedings of the {SIGCHI} Conference on Human
  Factors in Computing Systems}} (Vancouver, BC, Canada)
  \emph{(\bibinfo{series}{CHI '11})}. \bibinfo{publisher}{ACM},
  \bibinfo{address}{New York, NY, USA}, \bibinfo{pages}{1061--1070}.
\newblock
\showISBNx{9781450302289}
\urldef\tempurl%
\url{https://doi.org/10.1145/1978942.1979100}
\showDOI{\tempurl}


\bibitem[Gray(2016)]%
        {Gray2016-lq}
\bibfield{author}{\bibinfo{person}{Colin~M Gray}.}
  \bibinfo{year}{2016}\natexlab{}.
\newblock \showarticletitle{What is the Content of ''Design Thinking''? Design
  Heuristics as Conceptual Repertoire}.
\newblock \bibinfo{journal}{\emph{International Journal of Engineering
  Education}} \bibinfo{volume}{32}, \bibinfo{number}{3B}
  (\bibinfo{year}{2016}), \bibinfo{pages}{1349--1355}.
\newblock
\urldef\tempurl%
\url{http://www.ijee.ie/latestissues/Vol32-3B/05_ijee3220ns.pdf}
\showURL{%
\tempurl}


\bibitem[Gray(2020)]%
        {Gray2020-yu}
\bibfield{author}{\bibinfo{person}{Colin~M Gray}.}
  \bibinfo{year}{2020}\natexlab{}.
\newblock \showarticletitle{Markers of Quality in Design Precedent}.
\newblock \bibinfo{journal}{\emph{International Journal of Designs for
  Learning}} \bibinfo{volume}{11}, \bibinfo{number}{3} (\bibinfo{year}{2020}),
  \bibinfo{pages}{1--12}.
\newblock
\urldef\tempurl%
\url{https://doi.org/10.14434/ijdl.v11i3.31193}
\showDOI{\tempurl}


\bibitem[Gray and Boling(2017)]%
        {Gray2017-dx}
\bibfield{author}{\bibinfo{person}{Colin~M Gray} {and}
  \bibinfo{person}{Elizabeth Boling}.} \bibinfo{year}{2017}\natexlab{}.
\newblock \showarticletitle{Designers' Articulation and Activation of
  Instrumental Design Judgments in {Cross-Cultural} User Research}.
\newblock In \bibinfo{booktitle}{\emph{Analysing Design Thinking: Studies of
  {Cross-Cultural} {Co-Creation}}}, \bibfield{editor}{\bibinfo{person}{Bo~T
  Christensen}, \bibinfo{person}{Linden~J Ball}, {and} \bibinfo{person}{Kim
  Halskov}} (Eds.). \bibinfo{publisher}{CRC Press}, \bibinfo{address}{Boca
  Raton, FL}, \bibinfo{pages}{191--214}.
\newblock


\bibitem[Gray and Chivukula(2019)]%
        {Gray2019-ep}
\bibfield{author}{\bibinfo{person}{Colin~M Gray} {and}
  \bibinfo{person}{Shruthi~Sai Chivukula}.} \bibinfo{year}{2019}\natexlab{}.
\newblock \showarticletitle{Ethical Mediation in {UX} Practice}. In
  \bibinfo{booktitle}{\emph{Proceedings of the 2019 {CHI} Conference on Human
  Factors in Computing Systems}} (Glasgow, Scotland Uk)
  \emph{(\bibinfo{series}{CHI '19}, \bibinfo{number}{Paper 178})}.
  \bibinfo{publisher}{Association for Computing Machinery},
  \bibinfo{address}{New York, NY, USA}, \bibinfo{pages}{1--11}.
\newblock
\showISBNx{9781450359702}
\urldef\tempurl%
\url{https://doi.org/10.1145/3290605.3300408}
\showDOI{\tempurl}


\bibitem[Gray et~al\mbox{.}(2015)]%
        {Gray2015-qi}
\bibfield{author}{\bibinfo{person}{Colin~M Gray}, \bibinfo{person}{Cesur
  Dagli}, \bibinfo{person}{Muruvvet Demiral-Uzan}, \bibinfo{person}{Funda
  Ergulec}, \bibinfo{person}{Verily Tan}, \bibinfo{person}{Abdullah~A
  Altuwaijri}, \bibinfo{person}{Khendum Gyabak}, \bibinfo{person}{Megan
  Hilligoss}, \bibinfo{person}{Remzi Kizilboga}, \bibinfo{person}{Kei Tomita},
  {and} \bibinfo{person}{Elizabeth Boling}.} \bibinfo{year}{2015}\natexlab{}.
\newblock \showarticletitle{{Judgment and Instructional Design: How {ID}
  Practitioners Work In Practice}}.
\newblock \bibinfo{journal}{\emph{Performance Improvement Quarterly}}
  \bibinfo{volume}{28}, \bibinfo{number}{3} (\bibinfo{date}{Oct.}
  \bibinfo{year}{2015}), \bibinfo{pages}{25--49}.
\newblock
\showISSN{0898-5952}
\urldef\tempurl%
\url{https://doi.org/10.1002/piq.21198}
\showDOI{\tempurl}


\bibitem[Hayes(2014)]%
        {Hayes2014-mp}
\bibfield{author}{\bibinfo{person}{Gillian~R Hayes}.}
  \bibinfo{year}{2014}\natexlab{}.
\newblock \showarticletitle{Knowing by Doing: Action Research as an Approach to
  {HCI}}.
\newblock In \bibinfo{booktitle}{\emph{Ways of Knowing in {HCI}}},
  \bibfield{editor}{\bibinfo{person}{Judith~S Olson} {and}
  \bibinfo{person}{Wendy~A Kellogg}} (Eds.). \bibinfo{publisher}{Springer New
  York}, \bibinfo{address}{New York, NY}, \bibinfo{pages}{49--68}.
\newblock
\showISBNx{9781493903788}
\urldef\tempurl%
\url{https://doi.org/10.1007/978-1-4939-0378-8\_3}
\showDOI{\tempurl}


\bibitem[Hild{\'e}n et~al\mbox{.}(2017)]%
        {Hilden2017-jk}
\bibfield{author}{\bibinfo{person}{Elina Hild{\'e}n}, \bibinfo{person}{Jarno
  Ojala}, {and} \bibinfo{person}{Kaisa V{\"a}{\"a}n{\"a}nen}.}
  \bibinfo{year}{2017}\natexlab{}.
\newblock \showarticletitle{Development of Context Cards: A Bus-specific
  Ideation Tool for Co-design Workshops}. In
  \bibinfo{booktitle}{\emph{Proceedings of the 21st International Academic
  Mindtrek Conference}} (Tampere, Finland)
  \emph{(\bibinfo{series}{AcademicMindtrek '17})}. \bibinfo{publisher}{ACM},
  \bibinfo{address}{New York, NY, USA}, \bibinfo{pages}{137--146}.
\newblock
\showISBNx{9781450354264}
\urldef\tempurl%
\url{https://doi.org/10.1145/3131085.3131092}
\showDOI{\tempurl}


\bibitem[H{\"o}{\"o}k et~al\mbox{.}(2015)]%
        {Hook2015-tt}
\bibfield{author}{\bibinfo{person}{Kristina H{\"o}{\"o}k},
  \bibinfo{person}{Peter Dalsgaard}, \bibinfo{person}{Stuart Reeves},
  \bibinfo{person}{Jeffrey Bardzell}, \bibinfo{person}{Jonas L{\"o}wgren},
  \bibinfo{person}{Erik Stolterman}, {and} \bibinfo{person}{Yvonne Rogers}.}
  \bibinfo{year}{2015}\natexlab{}.
\newblock \showarticletitle{Knowledge Production in Interaction Design}. In
  \bibinfo{booktitle}{\emph{Proceedings of the 33rd Annual {ACM} Conference
  Extended Abstracts on Human Factors in Computing Systems - {CHI} {EA} '15}}.
  \bibinfo{publisher}{ACM Press}, \bibinfo{address}{New York, New York, USA},
  \bibinfo{pages}{2429--2432}.
\newblock
\showISBNx{9781450331463}
\urldef\tempurl%
\url{https://doi.org/10.1145/2702613.2702653}
\showDOI{\tempurl}


\bibitem[H{\"o}{\"o}k and L{\"o}wgren(2012)]%
        {Hook2012-dd}
\bibfield{author}{\bibinfo{person}{Kristina H{\"o}{\"o}k} {and}
  \bibinfo{person}{Jonas L{\"o}wgren}.} \bibinfo{year}{2012}\natexlab{}.
\newblock \showarticletitle{{Strong concepts}}.
\newblock \bibinfo{journal}{\emph{ACM transactions on computer-human
  interaction: a publication of the Association for Computing Machinery}}
  \bibinfo{volume}{19}, \bibinfo{number}{3} (\bibinfo{date}{Oct.}
  \bibinfo{year}{2012}), \bibinfo{pages}{1--18}.
\newblock
\showISSN{1073-0516}
\urldef\tempurl%
\url{https://doi.org/10.1145/2362364.2362371}
\showDOI{\tempurl}


\bibitem[Kleinsmann and Valkenburg(2008)]%
        {Kleinsmann2008-du}
\bibfield{author}{\bibinfo{person}{Maaike Kleinsmann} {and}
  \bibinfo{person}{Rianne Valkenburg}.} \bibinfo{year}{2008}\natexlab{}.
\newblock \showarticletitle{Barriers and enablers for creating shared
  understanding in co-design projects}.
\newblock \bibinfo{journal}{\emph{Design Studies}} \bibinfo{volume}{29},
  \bibinfo{number}{4} (\bibinfo{date}{July} \bibinfo{year}{2008}),
  \bibinfo{pages}{369--386}.
\newblock
\showISSN{0142-694X}
\urldef\tempurl%
\url{https://doi.org/10.1016/j.destud.2008.03.003}
\showDOI{\tempurl}


\bibitem[Kolari{\'c} et~al\mbox{.}(2020)]%
        {Kolaric2020-qc}
\bibfield{author}{\bibinfo{person}{Sini{\v s}a Kolari{\'c}},
  \bibinfo{person}{Jordan Beck}, {and} \bibinfo{person}{Erik Stolterman}.}
  \bibinfo{year}{2020}\natexlab{}.
\newblock \showarticletitle{On the Hierarchical Levels of Design Knowledge}.
\newblock \bibinfo{journal}{\emph{Proceedings of the Design Society: DESIGN
  Conference}}  \bibinfo{volume}{1} (\bibinfo{date}{May} \bibinfo{year}{2020}),
  \bibinfo{pages}{51--60}.
\newblock
\showISSN{2633-7762}
\urldef\tempurl%
\url{https://doi.org/10.1017/dsd.2020.330}
\showDOI{\tempurl}


\bibitem[Kronqvist and Salmi(2011)]%
        {Kronqvist2011-py}
\bibfield{author}{\bibinfo{person}{Juha Kronqvist} {and} \bibinfo{person}{Anna
  Salmi}.} \bibinfo{year}{2011}\natexlab{}.
\newblock \showarticletitle{Co-designing (with) Organizations:
  Human-centeredness, Participation and Embodiment in Organizational
  Development}. In \bibinfo{booktitle}{\emph{Proceedings of the 2011 Conference
  on Designing Pleasurable Products and Interfaces}} (Milano, Italy)
  \emph{(\bibinfo{series}{DPPI '11})}. \bibinfo{publisher}{ACM},
  \bibinfo{address}{New York, NY, USA}, \bibinfo{pages}{37:1--37:8}.
\newblock
\showISBNx{9781450312806}
\urldef\tempurl%
\url{https://doi.org/10.1145/2347504.2347544}
\showDOI{\tempurl}


\bibitem[Kuo et~al\mbox{.}(2019)]%
        {Kuo2019-uh}
\bibfield{author}{\bibinfo{person}{P~Y Kuo}, \bibinfo{person}{R Saran},
  \bibinfo{person}{M Argentina}, \bibinfo{person}{M Heung}, {and}
  \bibinfo{person}{{others}}.} \bibinfo{year}{2019}\natexlab{}.
\newblock \showarticletitle{Development of a Checklist for the Prevention of
  Intradialytic Hypotension in Hemodialysis Care: Design Considerations Based
  on Activity Theory}.
\newblock \bibinfo{journal}{\emph{Proceedings of the}} (\bibinfo{year}{2019}).
\newblock


\bibitem[Lawson(2004)]%
        {Lawson2004-wf}
\bibfield{author}{\bibinfo{person}{Bryan Lawson}.}
  \bibinfo{year}{2004}\natexlab{}.
\newblock \showarticletitle{{Schemata, gambits and precedent: some factors in
  design expertise}}.
\newblock \bibinfo{journal}{\emph{Design Studies}} \bibinfo{volume}{25},
  \bibinfo{number}{5} (\bibinfo{date}{Sept.} \bibinfo{year}{2004}),
  \bibinfo{pages}{443--457}.
\newblock
\showISSN{0142-694X}
\urldef\tempurl%
\url{https://doi.org/10.1016/j.destud.2004.05.001}
\showDOI{\tempurl}


\bibitem[Lee et~al\mbox{.}(2018)]%
        {Lee2018-ne}
\bibfield{author}{\bibinfo{person}{Jung-Joo Lee}, \bibinfo{person}{Miia
  Jaatinen}, \bibinfo{person}{Anna Salmi}, \bibinfo{person}{Tuuli
  Mattelm{\"a}ki}, \bibinfo{person}{Riitta Smeds}, {and} \bibinfo{person}{Mari
  Holopainen}.} \bibinfo{year}{2018}\natexlab{}.
\newblock \showarticletitle{Design choices framework for co-creation projects}.
\newblock \bibinfo{journal}{\emph{International Journal of Design}}
  \bibinfo{volume}{12}, \bibinfo{number}{2} (\bibinfo{year}{2018}).
\newblock


\bibitem[Lee(2008)]%
        {Lee2008-sn}
\bibfield{author}{\bibinfo{person}{Yanki Lee}.}
  \bibinfo{year}{2008}\natexlab{}.
\newblock \showarticletitle{Design participation tactics: the challenges and
  new roles for designers in the co-design process}.
\newblock \bibinfo{journal}{\emph{CoDesign}} \bibinfo{volume}{4},
  \bibinfo{number}{1} (\bibinfo{date}{March} \bibinfo{year}{2008}),
  \bibinfo{pages}{31--50}.
\newblock
\showISSN{1571-0882}
\urldef\tempurl%
\url{https://doi.org/10.1080/15710880701875613}
\showDOI{\tempurl}


\bibitem[Light and Akama(2012)]%
        {Light2012-zg}
\bibfield{author}{\bibinfo{person}{Ann Light} {and} \bibinfo{person}{Yoko
  Akama}.} \bibinfo{year}{2012}\natexlab{}.
\newblock \showarticletitle{The human touch: participatory practice and the
  role of facilitation in designing with communities}. In
  \bibinfo{booktitle}{\emph{Proceedings of the 12th Participatory Design
  Conference: Research {Papers-Volume} 1}}. \bibinfo{pages}{61--70}.
\newblock


\bibitem[Loi et~al\mbox{.}(2019)]%
        {Loi2019-tl}
\bibfield{author}{\bibinfo{person}{Daria Loi}, \bibinfo{person}{Christine~T
  Wolf}, \bibinfo{person}{Jeanette~L Blomberg}, \bibinfo{person}{Raphael Arar},
  {and} \bibinfo{person}{Margot Brereton}.} \bibinfo{year}{2019}\natexlab{}.
\newblock \showarticletitle{Co-designing {AI} Futures: Integrating {AI} Ethics,
  Social Computing, and Design}. In \bibinfo{booktitle}{\emph{Companion
  Publication of the 2019 on Designing Interactive Systems Conference 2019
  Companion}} (San Diego, CA, USA) \emph{(\bibinfo{series}{DIS '19
  Companion})}. \bibinfo{publisher}{ACM}, \bibinfo{address}{New York, NY, USA},
  \bibinfo{pages}{381--384}.
\newblock
\showISBNx{9781450362702}
\urldef\tempurl%
\url{https://doi.org/10.1145/3301019.3320000}
\showDOI{\tempurl}


\bibitem[L{\"o}wgren(2006)]%
        {Lowgren2006-vr}
\bibfield{author}{\bibinfo{person}{J L{\"o}wgren}.}
  \bibinfo{year}{2006}\natexlab{}.
\newblock \showarticletitle{{Articulating the use qualities of digital
  designs}}.
\newblock In \bibinfo{booktitle}{\emph{Aesthetic computing}},
  \bibfield{editor}{\bibinfo{person}{Paul~A Fishwick}} (Ed.).
  \bibinfo{publisher}{MIT Press}, \bibinfo{pages}{383--403}.
\newblock
\showISBNx{9780262062503}


\bibitem[L{\"o}wgren(2013)]%
        {Lowgren2013-db}
\bibfield{author}{\bibinfo{person}{Jonas L{\"o}wgren}.}
  \bibinfo{year}{2013}\natexlab{}.
\newblock \showarticletitle{{Annotated portfolios and other forms of
  intermediate-level knowledge}}.
\newblock \bibinfo{journal}{\emph{Interactions}} \bibinfo{volume}{20},
  \bibinfo{number}{1} (\bibinfo{date}{Jan.} \bibinfo{year}{2013}),
  \bibinfo{pages}{30--34}.
\newblock
\showISSN{1072-5520}
\urldef\tempurl%
\url{https://doi.org/10.1145/2405716.2405725}
\showDOI{\tempurl}


\bibitem[Madaio et~al\mbox{.}(2020)]%
        {Madaio2020-gu}
\bibfield{author}{\bibinfo{person}{Michael~A Madaio}, \bibinfo{person}{Luke
  Stark}, \bibinfo{person}{Jennifer Wortman~Vaughan}, {and}
  \bibinfo{person}{Hanna Wallach}.} \bibinfo{year}{2020}\natexlab{}.
\newblock \showarticletitle{{Co-Designing} Checklists to Understand
  Organizational Challenges and Opportunities around Fairness in {AI}}. In
  \bibinfo{booktitle}{\emph{Proceedings of the 2020 {CHI} Conference on Human
  Factors in Computing Systems}} (Honolulu, HI, USA)
  \emph{(\bibinfo{series}{CHI '20})}. \bibinfo{publisher}{Association for
  Computing Machinery}, \bibinfo{address}{New York, NY, USA},
  \bibinfo{pages}{1--14}.
\newblock
\showISBNx{9781450367080}
\urldef\tempurl%
\url{https://doi.org/10.1145/3313831.3376445}
\showDOI{\tempurl}


\bibitem[Manzini(2015)]%
        {Manzini2015-om}
\bibfield{author}{\bibinfo{person}{Ezio Manzini}.}
  \bibinfo{year}{2015}\natexlab{}.
\newblock \bibinfo{booktitle}{\emph{Design, When Everybody Designs: An
  Introduction to Design for Social Innovation}}.
\newblock \bibinfo{publisher}{MIT Press}.
\newblock
\showISBNx{9780262328647}
\urldef\tempurl%
\url{https://play.google.com/store/books/details?id=BDnqBgAAQBAJ}
\showURL{%
\tempurl}


\bibitem[Marttila and Botero(2013)]%
        {Marttila2013-ax}
\bibfield{author}{\bibinfo{person}{Sanna Marttila} {and}
  \bibinfo{person}{Andrea Botero}.} \bibinfo{year}{2013}\natexlab{}.
\newblock \showarticletitle{{The'Openness} Turn'in co-design. From usability,
  sociability and designability towards openness}.
\newblock \bibinfo{journal}{\emph{Smeds \& Irrmann (eds) CO-CREATE}}
  (\bibinfo{year}{2013}), \bibinfo{pages}{99--111}.
\newblock


\bibitem[McKercher(2020)]%
        {McKercher2020-xq}
\bibfield{author}{\bibinfo{person}{Kelly~Ann McKercher}.}
  \bibinfo{year}{2020}\natexlab{}.
\newblock \bibinfo{booktitle}{\emph{Beyond Sticky Notes: {Co-Design} for Real:
  Mindsets, Methods and Movements}}.
\newblock \bibinfo{publisher}{Beyond Sticky Notes}.
\newblock
\showISBNx{9780648787501}


\bibitem[Nelson and Stolterman(2012)]%
        {Nelson2012-ov}
\bibfield{author}{\bibinfo{person}{Harold~G Nelson} {and} \bibinfo{person}{Erik
  Stolterman}.} \bibinfo{year}{2012}\natexlab{}.
\newblock \bibinfo{booktitle}{\emph{The design way : Intentional change in an
  unpredictable world} (\bibinfo{edition}{2nd} ed.)}.
\newblock \bibinfo{publisher}{MIT Press}, \bibinfo{address}{Cambridge, MA}.
\newblock
\showISBNx{9780877783053}


\bibitem[Olivier and Wright(2015)]%
        {Olivier2015-dt}
\bibfield{author}{\bibinfo{person}{Patrick Olivier} {and}
  \bibinfo{person}{Peter Wright}.} \bibinfo{year}{2015}\natexlab{}.
\newblock \showarticletitle{Digital civics: taking a local turn}.
\newblock \bibinfo{journal}{\emph{Interactions}} \bibinfo{volume}{22},
  \bibinfo{number}{4} (\bibinfo{date}{June} \bibinfo{year}{2015}),
  \bibinfo{pages}{61--63}.
\newblock
\showISSN{1072-5520}
\urldef\tempurl%
\url{https://doi.org/10.1145/2776885}
\showDOI{\tempurl}


\bibitem[Sanders(2002)]%
        {Sanders2002-na}
\bibfield{author}{\bibinfo{person}{Elizabeth Sanders}.}
  \bibinfo{year}{2002}\natexlab{}.
\newblock \showarticletitle{From user-centered to participatory design
  approaches}.
\newblock In \bibinfo{booktitle}{\emph{Design and the social sciences}}.
  \bibinfo{publisher}{CRC Press}, \bibinfo{pages}{18--25}.
\newblock


\bibitem[Sanders and Stappers(2008)]%
        {Sanders2008-eq}
\bibfield{author}{\bibinfo{person}{Elizabeth Sanders} {and}
  \bibinfo{person}{Pieter~Jan Stappers}.} \bibinfo{year}{2008}\natexlab{}.
\newblock \showarticletitle{Co-creation and the new landscapes of design}.
\newblock \bibinfo{journal}{\emph{CoDesign}} \bibinfo{volume}{4},
  \bibinfo{number}{1} (\bibinfo{date}{March} \bibinfo{year}{2008}),
  \bibinfo{pages}{5--18}.
\newblock
\showISSN{1571-0882}
\urldef\tempurl%
\url{https://doi.org/10.1080/15710880701875068}
\showDOI{\tempurl}


\bibitem[Sanders and Stappers(2012)]%
        {Sanders2012-om}
\bibfield{author}{\bibinfo{person}{Elizabeth Sanders} {and}
  \bibinfo{person}{Pieter~Jan Stappers}.} \bibinfo{year}{2012}\natexlab{}.
\newblock \bibinfo{booktitle}{\emph{Convivial Toolbox: Generative Research for
  the Front End of Design}}.
\newblock \bibinfo{publisher}{BIS}.
\newblock
\showISBNx{9789063692841}


\bibitem[Sanders and Westerlund(2011)]%
        {Sanders2011-wm}
\bibfield{author}{\bibinfo{person}{Elizabeth B-N Sanders} {and}
  \bibinfo{person}{Bo Westerlund}.} \bibinfo{year}{2011}\natexlab{}.
\newblock \showarticletitle{Experiencing, Exploring and Experimenting in and
  with {Co-Design} Spaces}.
\newblock \bibinfo{journal}{\emph{Nordes}} \bibinfo{volume}{0},
  \bibinfo{number}{4} (\bibinfo{date}{March} \bibinfo{year}{2011}).
\newblock
\showISSN{1604-9705, 1604-9705}


\bibitem[Sch{\"o}n(1983)]%
        {Schon1983-dl}
\bibfield{author}{\bibinfo{person}{Donald~A Sch{\"o}n}.}
  \bibinfo{year}{1983}\natexlab{}.
\newblock \bibinfo{booktitle}{\emph{The reflective practitioner: How
  professionals think in action}}.
\newblock \bibinfo{publisher}{Basic Books}, \bibinfo{address}{New York, NY}.
\newblock
\showISBNx{9781351883153}


\bibitem[Sch{\"o}n(1990)]%
        {Schon1990-by}
\bibfield{author}{\bibinfo{person}{Donald~A Sch{\"o}n}.}
  \bibinfo{year}{1990}\natexlab{}.
\newblock \showarticletitle{The design process}.
\newblock In \bibinfo{booktitle}{\emph{Varieties of thinking: Essays from
  Harvard's philosophy of education research center}},
  \bibfield{editor}{\bibinfo{person}{V~A Howard}} (Ed.).
  \bibinfo{publisher}{Routledge}, \bibinfo{pages}{111--141}.
\newblock


\bibitem[Schuler and Namioka(1993)]%
        {Schuler1993-ut}
\bibfield{editor}{\bibinfo{person}{Douglas Schuler} {and} \bibinfo{person}{Aki
  Namioka}} (Eds.). \bibinfo{year}{1993}\natexlab{}.
\newblock \bibinfo{booktitle}{\emph{Participatory Design: Principles and
  Practices}}.
\newblock \bibinfo{publisher}{L. Erlbaum Associates Inc.},
  \bibinfo{address}{Hillsdale, NJ, USA}.
\newblock
\showISBNx{9780805809510}


\bibitem[Shakespeare and Marlow(2014)]%
        {Shakespeare2014-rf}
\bibfield{author}{\bibinfo{person}{Pauline Shakespeare} {and}
  \bibinfo{person}{Oliver Marlow}.} \bibinfo{year}{2014}\natexlab{}.
\newblock \bibinfo{title}{Whittington Hospital Pharmacy}.
\newblock
  \bibinfo{howpublished}{\url{http://www.designforeurope.eu/case-study/whittington-hospital-pharmacy}}.
\newblock
\newblock
\shownote{Accessed: 2020-2-21}.


\bibitem[Steen(2011)]%
        {Steen2011-qo}
\bibfield{author}{\bibinfo{person}{Marc Steen}.}
  \bibinfo{year}{2011}\natexlab{}.
\newblock \showarticletitle{Tensions in human-centred design}.
\newblock \bibinfo{journal}{\emph{CoDesign}} \bibinfo{volume}{7},
  \bibinfo{number}{1} (\bibinfo{date}{March} \bibinfo{year}{2011}),
  \bibinfo{pages}{45--60}.
\newblock
\showISSN{1571-0882}
\urldef\tempurl%
\url{https://doi.org/10.1080/15710882.2011.563314}
\showDOI{\tempurl}


\bibitem[Steen(2013)]%
        {Steen2013-is}
\bibfield{author}{\bibinfo{person}{Marc Steen}.}
  \bibinfo{year}{2013}\natexlab{}.
\newblock \showarticletitle{{Co-Design} as a Process of Joint Inquiry and
  Imagination}.
\newblock \bibinfo{journal}{\emph{Design Issues}} \bibinfo{volume}{29},
  \bibinfo{number}{2} (\bibinfo{date}{April} \bibinfo{year}{2013}),
  \bibinfo{pages}{16--28}.
\newblock
\showISSN{0747-9360}
\urldef\tempurl%
\url{https://doi.org/10.1162/DESI\_a\_00207}
\showDOI{\tempurl}


\bibitem[Steen et~al\mbox{.}(2011)]%
        {Steen2011-ue}
\bibfield{author}{\bibinfo{person}{Marc Steen}, \bibinfo{person}{Menno
  Manschot}, {and} \bibinfo{person}{Nicole De~Koning}.}
  \bibinfo{year}{2011}\natexlab{}.
\newblock \showarticletitle{Benefits of co-design in service design projects}.
\newblock \bibinfo{journal}{\emph{International Journal of Design}}
  \bibinfo{volume}{5}, \bibinfo{number}{2} (\bibinfo{year}{2011}).
\newblock


\bibitem[Stigberg(2017)]%
        {Stigberg2017-ni}
\bibfield{author}{\bibinfo{person}{Susanne~Koch Stigberg}.}
  \bibinfo{year}{2017}\natexlab{}.
\newblock \showarticletitle{Mobile Hand Gesture Toolkit: {Co-Designing} Mobile
  Interaction Interfaces}. In \bibinfo{booktitle}{\emph{Proceedings of the 2017
  {ACM} Conference Companion Publication on Designing Interactive Systems}}
  (Edinburgh, United Kingdom) \emph{(\bibinfo{series}{DIS '17 Companion})}.
  \bibinfo{publisher}{ACM}, \bibinfo{address}{New York, NY, USA},
  \bibinfo{pages}{161--166}.
\newblock
\showISBNx{9781450349918}
\urldef\tempurl%
\url{https://doi.org/10.1145/3064857.3079138}
\showDOI{\tempurl}


\bibitem[Stolterman(2008)]%
        {Stolterman2008-ho}
\bibfield{author}{\bibinfo{person}{E Stolterman}.}
  \bibinfo{year}{2008}\natexlab{}.
\newblock \showarticletitle{{The nature of design practice and implications for
  interaction design research}}.
\newblock \bibinfo{journal}{\emph{International Journal of Design}}
  \bibinfo{volume}{2}, \bibinfo{number}{1} (\bibinfo{date}{Jan.}
  \bibinfo{year}{2008}), \bibinfo{pages}{55--65}.
\newblock
\showISSN{0950-1991}
\urldef\tempurl%
\url{https://doi.org/10.1016/j.phymed.2007.09.005}
\showDOI{\tempurl}


\bibitem[Tidwell(2010)]%
        {Tidwell2010-tq}
\bibfield{author}{\bibinfo{person}{Jenifer Tidwell}.}
  \bibinfo{year}{2010}\natexlab{}.
\newblock \bibinfo{booktitle}{\emph{Designing Interfaces: Patterns for
  Effective Interaction Design}}.
\newblock \bibinfo{publisher}{O'Reilly Media, Inc.}
\newblock
\showISBNx{9781449302733}
\urldef\tempurl%
\url{https://market.android.com/details?id=book-5gvOU9X0fu0C}
\showURL{%
\tempurl}


\bibitem[Vaajakallio and Mattelm{\"a}ki(2014)]%
        {Vaajakallio2014-bu}
\bibfield{author}{\bibinfo{person}{Kirsikka Vaajakallio} {and}
  \bibinfo{person}{Tuuli Mattelm{\"a}ki}.} \bibinfo{year}{2014}\natexlab{}.
\newblock \showarticletitle{Design games in codesign: as a tool, a mindset and
  a structure}.
\newblock \bibinfo{journal}{\emph{CoDesign}} \bibinfo{volume}{10},
  \bibinfo{number}{1} (\bibinfo{date}{Jan.} \bibinfo{year}{2014}),
  \bibinfo{pages}{63--77}.
\newblock
\showISSN{1571-0882}
\urldef\tempurl%
\url{https://doi.org/10.1080/15710882.2014.881886}
\showDOI{\tempurl}


\bibitem[Verbeek(2006)]%
        {Verbeek2006-qz}
\bibfield{author}{\bibinfo{person}{Peter-Paul Verbeek}.}
  \bibinfo{year}{2006}\natexlab{}.
\newblock \showarticletitle{{Materializing Morality: Design Ethics and
  Technological Mediation}}.
\newblock In \bibinfo{booktitle}{\emph{Science, Technology \& Human Values}}.
  Vol.~\bibinfo{volume}{31}. \bibinfo{pages}{361--380}.
\newblock
\showISSN{0162-2439}
\urldef\tempurl%
\url{https://doi.org/10.1177/0162243905285847}
\showDOI{\tempurl}


\bibitem[Visser et~al\mbox{.}(2005)]%
        {Visser2005-kl}
\bibfield{author}{\bibinfo{person}{Froukje~Sleeswijk Visser},
  \bibinfo{person}{Pieter~Jan Stappers}, \bibinfo{person}{Remko van~der Lugt},
  {and} \bibinfo{person}{Elizabeth Sanders}.} \bibinfo{year}{2005}\natexlab{}.
\newblock \showarticletitle{Contextmapping: experiences from practice}.
\newblock \bibinfo{journal}{\emph{CoDesign}} \bibinfo{volume}{1},
  \bibinfo{number}{2} (\bibinfo{date}{April} \bibinfo{year}{2005}),
  \bibinfo{pages}{119--149}.
\newblock
\showISSN{1571-0882}
\urldef\tempurl%
\url{https://doi.org/10.1080/15710880500135987}
\showDOI{\tempurl}


\bibitem[Yoo et~al\mbox{.}(2013)]%
        {Yoo2013-qu}
\bibfield{author}{\bibinfo{person}{Daisy Yoo}, \bibinfo{person}{Alina
  Huldtgren}, \bibinfo{person}{Jill~Palzkill Woelfer}, \bibinfo{person}{David~G
  Hendry}, {and} \bibinfo{person}{Batya Friedman}.}
  \bibinfo{year}{2013}\natexlab{}.
\newblock \showarticletitle{A value sensitive action-reflection model: evolving
  a co-design space with stakeholder and designer prompts}. In
  \bibinfo{booktitle}{\emph{Proceedings of the {SIGCHI} conference on human
  factors in computing systems}}. \bibinfo{pages}{419--428}.
\newblock


\bibitem[Yoo et~al\mbox{.}(2010)]%
        {Yoo2010-gy}
\bibfield{author}{\bibinfo{person}{Daisy Yoo}, \bibinfo{person}{John
  Zimmerman}, \bibinfo{person}{Aaron Steinfeld}, {and} \bibinfo{person}{Anthony
  Tomasic}.} \bibinfo{year}{2010}\natexlab{}.
\newblock \showarticletitle{Understanding the space for co-design in riders'
  interactions with a transit service}. In
  \bibinfo{booktitle}{\emph{Proceedings of the {SIGCHI} Conference on Human
  Factors in Computing Systems}}. \bibinfo{pages}{1797--1806}.
\newblock


\bibitem[Zimmerman and Forlizzi(2014)]%
        {Zimmerman2014-zx}
\bibfield{author}{\bibinfo{person}{John Zimmerman} {and} \bibinfo{person}{Jodi
  Forlizzi}.} \bibinfo{year}{2014}\natexlab{}.
\newblock \showarticletitle{Research Through Design in {HCI}}.
\newblock In \bibinfo{booktitle}{\emph{Ways of Knowing in {HCI}}},
  \bibfield{editor}{\bibinfo{person}{Judith~S Olson} {and}
  \bibinfo{person}{Wendy~A Kellogg}} (Eds.). \bibinfo{publisher}{Springer New
  York}, \bibinfo{address}{New York, NY}, \bibinfo{pages}{167--189}.
\newblock
\showISBNx{9781493903788}
\urldef\tempurl%
\url{https://doi.org/10.1007/978-1-4939-0378-8\_8}
\showDOI{\tempurl}


\bibitem[Zimmerman et~al\mbox{.}(2007)]%
        {Zimmerman2007-qi}
\bibfield{author}{\bibinfo{person}{John Zimmerman}, \bibinfo{person}{Jodi
  Forlizzi}, {and} \bibinfo{person}{Shelley Evenson}.}
  \bibinfo{year}{2007}\natexlab{}.
\newblock \showarticletitle{Research through design as a method for interaction
  design research in {HCI}}. In \bibinfo{booktitle}{\emph{Proceedings of the
  {SIGCHI} conference on Human factors in computing systems - {CHI} '07}}.
  \bibinfo{publisher}{ACM Press}, \bibinfo{address}{New York, New York, USA},
  \bibinfo{pages}{493--502}.
\newblock
\showISBNx{9781595935939}
\urldef\tempurl%
\url{https://doi.org/10.1145/1240624.1240704}
\showDOI{\tempurl}


\bibitem[Zimmerman et~al\mbox{.}(2010)]%
        {Zimmerman2010-lm}
\bibfield{author}{\bibinfo{person}{John Zimmerman}, \bibinfo{person}{Erik
  Stolterman}, {and} \bibinfo{person}{Jodi Forlizzi}.}
  \bibinfo{year}{2010}\natexlab{}.
\newblock \showarticletitle{An analysis and critique of Research through
  Design: towards a formalization of a research approach}. In
  \bibinfo{booktitle}{\emph{Proceedings of the 8th {ACM} Conference on
  Designing Interactive Systems}} (Aarhus, Denmark) \emph{(\bibinfo{series}{DIS
  '10})}. \bibinfo{publisher}{Association for Computing Machinery},
  \bibinfo{address}{New York, NY, USA}, \bibinfo{pages}{310--319}.
\newblock
\showISBNx{9781450301039}
\urldef\tempurl%
\url{https://doi.org/10.1145/1858171.1858228}
\showDOI{\tempurl}


\end{thebibliography}


\end{document}